\documentclass[12pt,preprint]{aastex}

\newcommand {\Lya}    {Ly$\alpha$}   %  Lyalpha
\newcommand {\Lyb}    {Ly$\beta$}      %  Lybeta

\newcommand {\HH}     {H$_2$}        %  H2
\newcommand {\HI}     {\ion{H}{1}}   %  HI
\newcommand {\HII}    {\ion{H}{2}}   %  HII
\newcommand {\OI}      {\ion{O}{1}}   %  OI
\newcommand {\OII}     {\ion{O}{2}}   %  OII
\newcommand {\OIV}    {\ion{O}{4}}   %  OIV
\newcommand {\OVI}    {\ion{O}{6}}   %  OVI

\newcommand {\CII}    {\ion{C}{2}}
\newcommand {\CIII}   {\ion{C}{3}}  
\newcommand {\CIV}    {\ion{C}{4}}

\newcommand {\NI}     {\ion{N}{1}}
\newcommand {\NV}    {\ion{N}{5}}

\newcommand {\SiII}    {\ion{Si}{2}}
\newcommand {\SiIII}   {\ion{Si}{3}}
\newcommand {\SiIV}  {\ion{Si}{4}}

\newcommand {\FeII}   {\ion{Fe}{2}}

\newcommand {\AlII}   {\ion{Al}{2}}
\newcommand {\SII}    {\ion{S}{2}}
\newcommand {\PII}    {\ion{P}{2}}
\newcommand {\NiII}    {\ion{Ni}{2}}

\newcommand {\kms}    {km~s$^{-1}$}
\newcommand {\NHI}    {$N_{\rm HI}$}

\newcommand {\FUSE}  {{\it FUSE}} 
\newcommand {\HST}  {{\it HST}}

\newcommand {\etal}  {et~al.} 
\newcommand {\cd}    {cm$^{-2}$}  

\slugcomment{Accepted for ApJ, Vol 738 (Sept. 1, 2011)}
\shorttitle{short title}
\shortauthors{author \etal}

\begin{document}

\title{Hubble-COS Observations of Galactic High-Velocity Clouds:  \\
   Four AGN Sight Lines through Complex~C
\footnote{Based on observations made with the NASA/ESA {\it Hubble Space Telescope}, obtained from the
data archive at the Space Telescope Science Institute. STScI is operated by the Association of Universities for
 Research in Astronomy, Inc. under NASA contract NAS5-26555.} }
\author{J. Michael Shull, Matthew Stevans, Charles Danforth, Steven V. Penton \\
CASA, Department of Astrophysical and Planetary Sciences, University of Colorado, 389-UCB, Boulder, CO 80309} 

\author{Felix J.\ Lockman, National Radio Astronomy Observatory, Green Bank, WV 29444; 
Nahum Arav, Department of Physics, Virginia Tech, Blacksburg, VA 24061 } 

\email{michael.shull@colorado.edu, matthew.stevans@colorado.edu, charles.danforth@colorado.edu, steven.penton@colorado.edu, jlockman@nrao.edu, arav@vt.edu}

\begin{abstract}

We report ultraviolet spectra of Galactic high-velocity clouds (HVCs) in Complex C, taken by the Cosmic 
Origins Spectrograph (COS) on the {\it Hubble Space Telescope} (\HST), together with new 21-cm spectra
from the Green Bank Telescope.   The wide spectral coverage and higher S/N, compared to previous HST
spectra, provide better velocity definition of the HVC absorption, additional ionization species (including
high ions), and improved abundances in this halo gas.
Complex C has a metallicity of 10--30\% solar and a wide range of ions, suggesting dynamical and thermal 
interactions with hot gas in the Galactic halo.  Spectra in the COS medium-resolution G130M (1133--1468~\AA) 
and G160M (1383--1796~\AA) gratings detect ultraviolet absorption
lines from 8 elements in low ionization stages (\OI, \NI, \CII, \SII, \SiII, \AlII,  \FeII, \PII) and 3 elements in 
intermediate and high-ionization states (\SiIII, \SiIV, \CIV, \NV).    Our four AGN sight lines toward Mrk~817, 
Mrk~290, Mrk~876, and PG~1259+593 have high-velocity  \HI\ and \OVI\ column densities,
$\log N_{\rm HI} =$ 19.39--20.05 and $\log N_{\rm OVI} =$ 13.58--14.10, with substantial amounts of 
kinematically associated photoionized gas.   The high-ion abundance ratios are consistent with cooling 
interfaces between photoionized and collisionally ionized gas:  N(\CIV)/N(\OVI) $\approx$ 0.3--0.5, 
N(\SiIV)/N(\OVI)  $\approx$ 0.05--0.11, N(\NV)/N(\OVI) $\approx$ 0.07--0.13, and 
N(\SiIV)/N(\SiIII) $\approx 0.2$.  

\end{abstract}


\newpage

%%%%   Section 1 -- Introduction   %%%%%%%%%%%%%%%%%%%%%%%%

\section{Introduction}

Absorption spectra in the ultraviolet (UV) provide sensitive diagnostics of conditions in the 
halo of the Milky Way, where competing process of accretion and outflow determine
 the evolution of the Galaxy.   Recent measurements (Shull \etal\ 2009) provide considerable 
 insight into the infall of low-metallicity gas onto the disk, an ongoing process that can account 
 for the observed stellar metallicities, star formation rates, and mass-metallicity relations 
(Pagel 1994; Gilmore 2001; Tremonti et al.\ 2004).  
Galactic high-velocity H~I clouds (HVCs) are plausible candidates for this fresh material.  
However, as long as they were only observed in the 21 cm line, their overall properties 
remained somewhat obscure.  In recent years, sensitive UV and optical spectroscopy have
revitalized this field, providing metallicities (10-20\% solar) and reliable distance measurements 
to several prominent clouds such as Complex C, Complex M, and the Magellanic Stream.   

The infall of low-metallicity gas onto the Milky~Way is a crucial component of most models 
of Galactic formation and evolution (Gibson \etal\ 2001) required to explain the metallicity 
distribution of nearby G- and K-dwarfs, the so-called ``G-dwarf problem'' (Pagel 1994).
Infall models are attractive because Galactic disk formation is believed to occur
by the gradual accretion of pristine or partially processed material into the interstellar
medium (ISM).   The metallicity of the 
initial reservoir of gas is enriched by ejecta from star formation and mixing with infalling 
low-metallicity gas, possibly through ``cold-mode accretion" (Dekel \& Birnboim 2006; 
Kere\v{s} \etal\ 2009).  
This process continues to the present day, regulated in a manner that produces the local 
G-dwarf metallicity distribution and avoids the overproduction of metal-poor disk stars.  
The infall of gas from the intergalactic medium (IGM) and low Galactic halo also places 
a chemical imprint on mass--metallicity relations (Erb \etal\ 2006).  

One likely manifestation of infall from the halo into the disk may have been observed in the 
system of Galactic high-velocity clouds (HVCs).  These HVCs were defined as neutral 
hydrogen clouds moving at velocities incompatible with differential Galactic rotation
(Wakker \& van Woerden 1997).  First discovered in 21-cm emission, Galactic HVCs have
become even more interesting when observed in UV absorption lines of heavy elements
(Wakker et al.\ 1999; Gibson \etal\ 2000, 2001; Richter \etal\ 2001; Sembach \etal\ 1999, 2003; 
Collins, Shull, \& Giroux  2003, 2004, 2005, 2007, 2009; Fox et al.\ 2004, 2006).  
Hereafter, we denote the Collins \etal\ papers as CSG03, CSG07, CSG09, etc.    
The UV data also demonstrate that HVCs are more extended on the sky at lower total 
hydrogen column densities.  Their ``extended atmospheres" often contain more ionized
gas than neutral gas, as seen in their H$\alpha$ emission (Tufte, Reynolds, \& Haffner 1998). 

The UV absorption-line surveys find that a greater fraction of the 
high-latitude sky is covered with infalling ionized gas than would have been 
suspected from 21 cm data.   In 21-cm emission, the HVC sky-covering factor is 37\% 
down to column densities N$_{\rm HI} \ga 8 \times 10^{17}$ \cd\  at the $4\sigma$ level
(Lockman \etal\ 2002).   In surveys using the more sensitive UV absorption lines, 
the sky coverage is much higher:   $\ga 60$\% in \OVI\ $\lambda1031.9$ (Sembach \etal\ 
2003) and $81 \pm 5$\% in \SiIII\ $\lambda1206.5$ (Shull \etal\ 2009).    
Owing to its large oscillator strength, the \SiIII\ absorption line is  the best probe of ionized 
HVCs, typically 4--5 times stronger than \OVI.  From our \HST/STIS survey of high-velocity 
\SiIII\ (Shull et al.\ 2009; CSG09), we infer that the low Galactic halo is enveloped by a sheath 
of ionized, low-metallicity gas, which can provide a substantial cooling inflow 
($\sim1~M_{\odot}$ yr$^{-1}$) to help replenish star formation in the Galactic disk,
estimated at 2--4 $M_{\odot}$ yr$^{-1}$ (Diehl \etal\ 2006; Robitaille \& Whitney 2010).  

Measuring the column densities, metallicities, and ionization conditions in HVCs are key 
steps in elucidating their importance in Galactic evolution.  These parameters are best 
measured in the UV, where a rich variety of elements and ion stages is accessible through 
their resonance absorption lines,  which are sensitive to column densities well below that 
detectable in 21-cm emission.   For example, \SiIII\ $\lambda 1206.5$ is easily detectable 
with \HST\  at column densities N$_{\rm SiIII} \ga 10^{12}$~cm$^{-2}$, corresponding to total 
hydrogen column densities N$_H \ga (3 \times 10^{16}$ cm$^{-2})(Z_{\odot}/Z)$, scaling 
inversely with metallicity.  The high-quality data achievable with COS (signal-to-noise 
S/N $\ga 30$) provide much better definition of
the velocity extent of these HVCs, many of which do not exhibit narrow Gaussian absorption 
profiles.  The ions commonly accessible to the COS G130/G160M gratings  
include  \OI, \NI, \NV, \CII, \CIV, \SiII, \SiIII, \SiIV, \SII, \AlII, \FeII, \NiII, and \PII.  
 
In this paper, we present high-quality, far-UV spectroscopic observations of absorption
along four sight lines passing through Complex~C.  One of the most prominent HVCs,
Complex~C extends over Galactic longitudes from $\ell \approx 30^{\circ}$ to $150^{\circ}$ in 
the Northern Galactic hemisphere (see maps by Wakker 2001, CSG03, Fox \etal\ 2004).  
Multiple sight lines pierce Complex C, with UV-determined metallicities 
ranging from 10--30\% solar (Wakker \etal\ 1999; CSG03, CSG07).  
The recent distance estimate of $d = 10 \pm 2.5$ kpc (Wakker \etal\ 2007; Thom et al.\ 2008) 
confirms that it has a substantial mass ($\sim 10^7~M_{\odot}$).   As Complex~C falls into the 
Galactic disk over the next 50--100 Myr, it will deliver an average mass inflow of
$\sim0.1~M_{\odot}$~yr$^{-1}$, some 5\% of the disk replenishment rate for star formation.

Our spectral data were obtained with the moderate-resolution ($R \approx 18,000$) gratings, 
G130M and G160M, on the Cosmic Origins Spectrograph (COS) on \HST\  (Green \etal\ 2011;  
Osterman \etal\ 2011).   The typical COS wavelength coverage is from 1133--1796~\AA, although 
individual spectra extend  slightly outside this range:  
Mrk 817 (1134.5--1796.1 \AA),  Mrk 876 (1135.4--1795.3 \AA), Mrk 290 (1134.1--1796.2 \AA),
PG\,1259+593 (1133.9--1796.1 \AA).  
These UV spectra demonstrate COS capabilities for detecting HVCs along 
sight lines to background AGN, with several improvements over previous studies.  
First, the much greater far-UV throughput of the COS gratings provides higher signal-to-noise
(S/N) and better photometric accuracy.  Second, the low background of the COS detectors allows 
us to characterize the zero flux levels, important for measurements of column densities of mildly
saturated absorption lines.  Third, the combination of G130M/G160M gratings offers
broad wavelength coverage and access to numerous ion species and resonant lines not covered 
previously, with either the Goddard High Resolution Spectrograph (GHRS) or  \FUSE.   
 In particular, our COS spectra measure species (\AlII, \CIV, \SiIV) and transitions (\OI, \SiII, \FeII, \PII)  
 not typically measured or reported in GHRS data.  Detecting  HVCs in multiple ionization stages 
 of the same element such as silicon (\SiII, \SiIII, \SiIV) provides 
diagnostics of the ionization conditions and metallicity of the HVCs, while higher ionization stages 
(\CIV, \NV, \OVI, \SiIII, \SiIV) are useful in separating the contribution of hot collisionally ionized gas 
from warm photoionized gas.  

In \S2 we discuss the observations and data reduction techniques for the COS gratings 
(G130M and G160M) and 21-cm spectra from the NRAO\footnote{The National 
Radio Astronomy Observatory is a facility of the National Science Foundation, operated under a 
cooperative agreement with Associated Universities, Inc.} Green Bank Telescope (GBT).  
In \S3 we display the data and describe our analysis.  
In \S4 we summarize our observations and their implications for Complex~C (metallicity,
ionization state, and velocity structure).

%%%%%%%%%   Section 2  %%%%%%%%%%%%%
%%%%%

\section{Observations of Complex C}

In this section, we describe the HST/COS ultraviolet spectra of four AGN sight lines (Table 1)
passing through Complex C, with absorption typically appearing at LSR velocities 
$V_{\rm LSR} \approx -160$ to $-90$ \kms.  Figure 1 illustrates the locations of our  
background AGN targets relative to the 21-cm emission from the Leiden-Argentine-Bonn (LAB)
survey (Kalberla \etal\ 2005) with angular resolution $\sim0.6^{\circ}$, spectral resolution 1.3 \kms, 
and 70--90 mK rms noise in brightness-temperature ($T_b$).  These four sight lines exhibit a wide 
range in \HI\ column densities from $\log N_{\rm HI} = 19.39$ to 20.05 (Table 2).  Figure 2 shows 
the combined G130M/G160M ultraviolet spectra from COS, over the wavelength range 1135--1796~\AA.  
New 21-cm spectra taken at the GBT are shown in Figure 3.

%%%%%%%%%%%   New Figure 1   %%%%%%%%%%%

\begin{figure}
   \epsscale{1.05} 
   \plotone{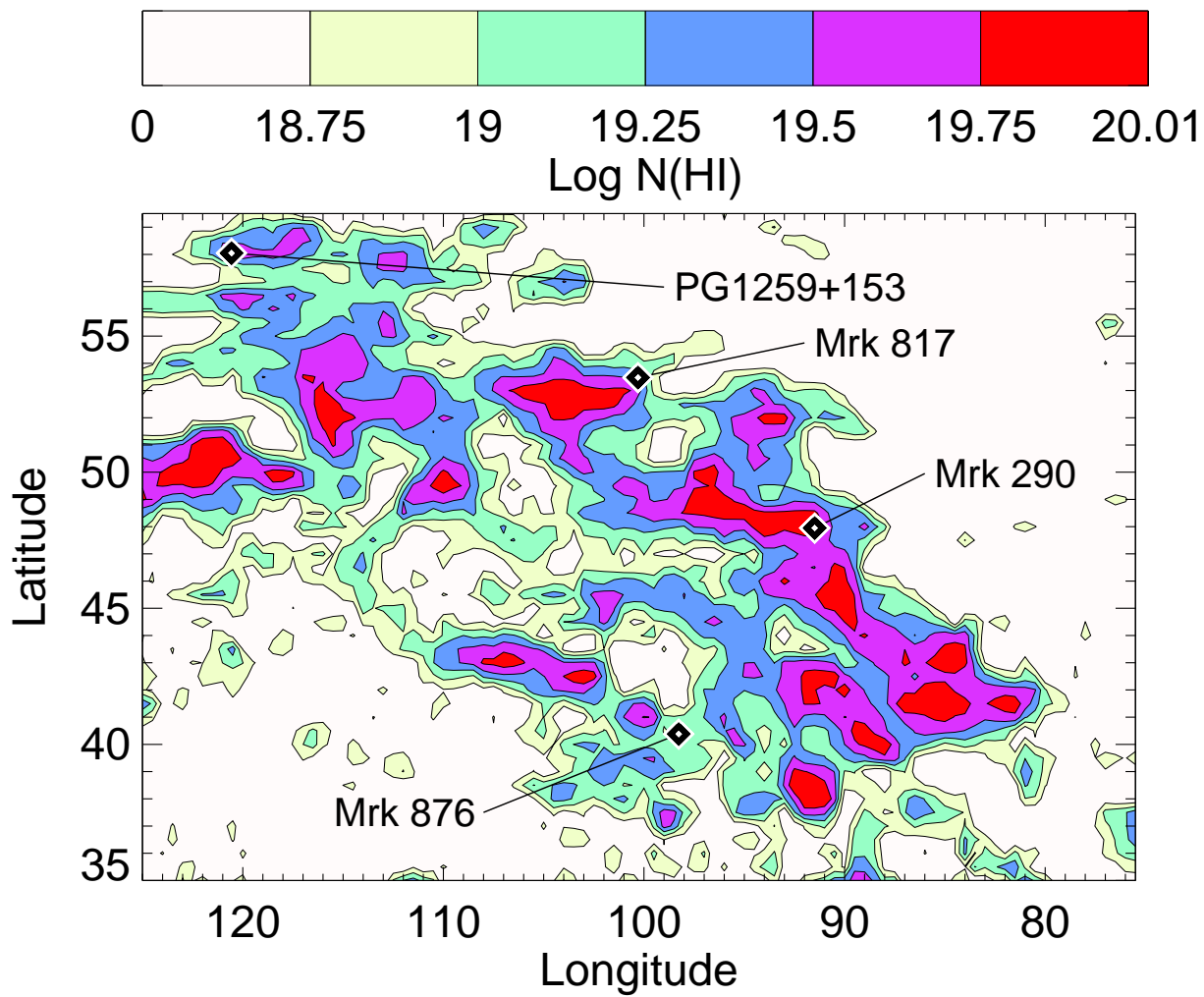}    
   \caption{Map of our 4 AGN targets overlaid on the contour-map of  21-cm emission
   from the Leiden-Argentine-Bonn (LAB) survey (Kalberla \etal\ 2005) with $\sim0.6^{\circ}$ 
   angular resolution on a $0.5^{\circ}$ grid in $\ell$ and $b$.  
   Emission is shown over Galactic coordinates between  $\ell = 75-125^{\circ}$ 
   and $b = 35-60^{\circ}$ and over Complex-C velocities between $-210$ and $-95$ \kms. 
   At the 10 kpc distance of Complex~C, 1 degree corresponds to 175 pc. }
  
\end{figure} 

%%%%%%%%%%%%%%%%%%%%%%%%%%%%%%%%%%%%%%%%%%%%%%%%

%%%%%%%%%%%   New Figure 2   %%%%%%%%%%%

\begin{figure}
   \epsscale{1.0} 
   \plotone{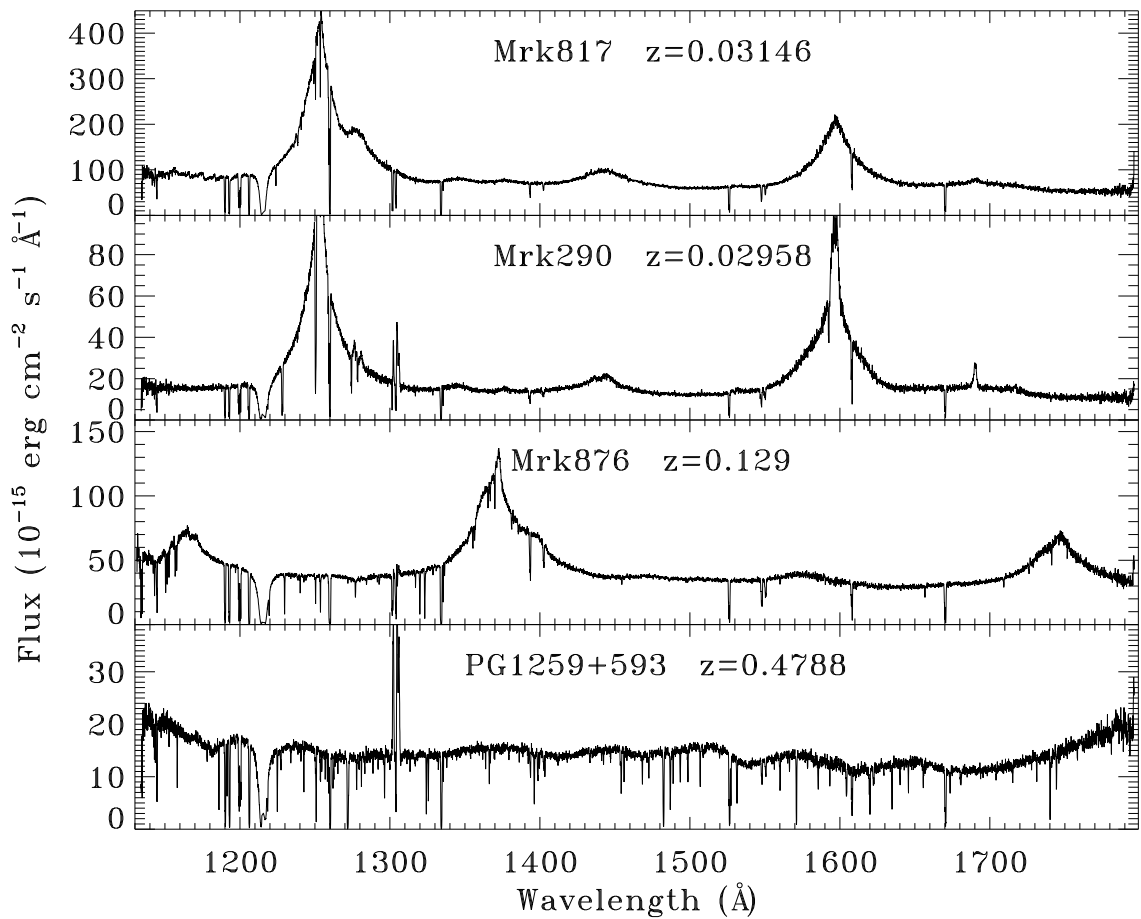}    
   \caption{Combined G130M/G160M spectra from HST/COS of the four AGN targets behind
   Complex C over the wavalength range 1135--1796 \AA.    Note the prominent broad emission
   lines, \Lya\ $\lambda 1216$, \SiIV/\OIV] $\lambda 1400$, \CIV\ $\lambda 1549$, together
   with numerous narrow absorption lines from the ISM and IGM.  
  Fluxes are in units of $10^{-15}$ erg~cm$^{-2}$ s$^{-1}$ \AA$^{-1}$.  
      }
\end{figure} 

%%%%%%%%%%%%%%%%%%%%%%%%%%%%%%%%%%%%%%%%%%%%%%%%

%%%%%%%%%%%   New Figure 3   %%%%%%%%%%%

\begin{figure}
   \epsscale{0.8} 
   \plotone{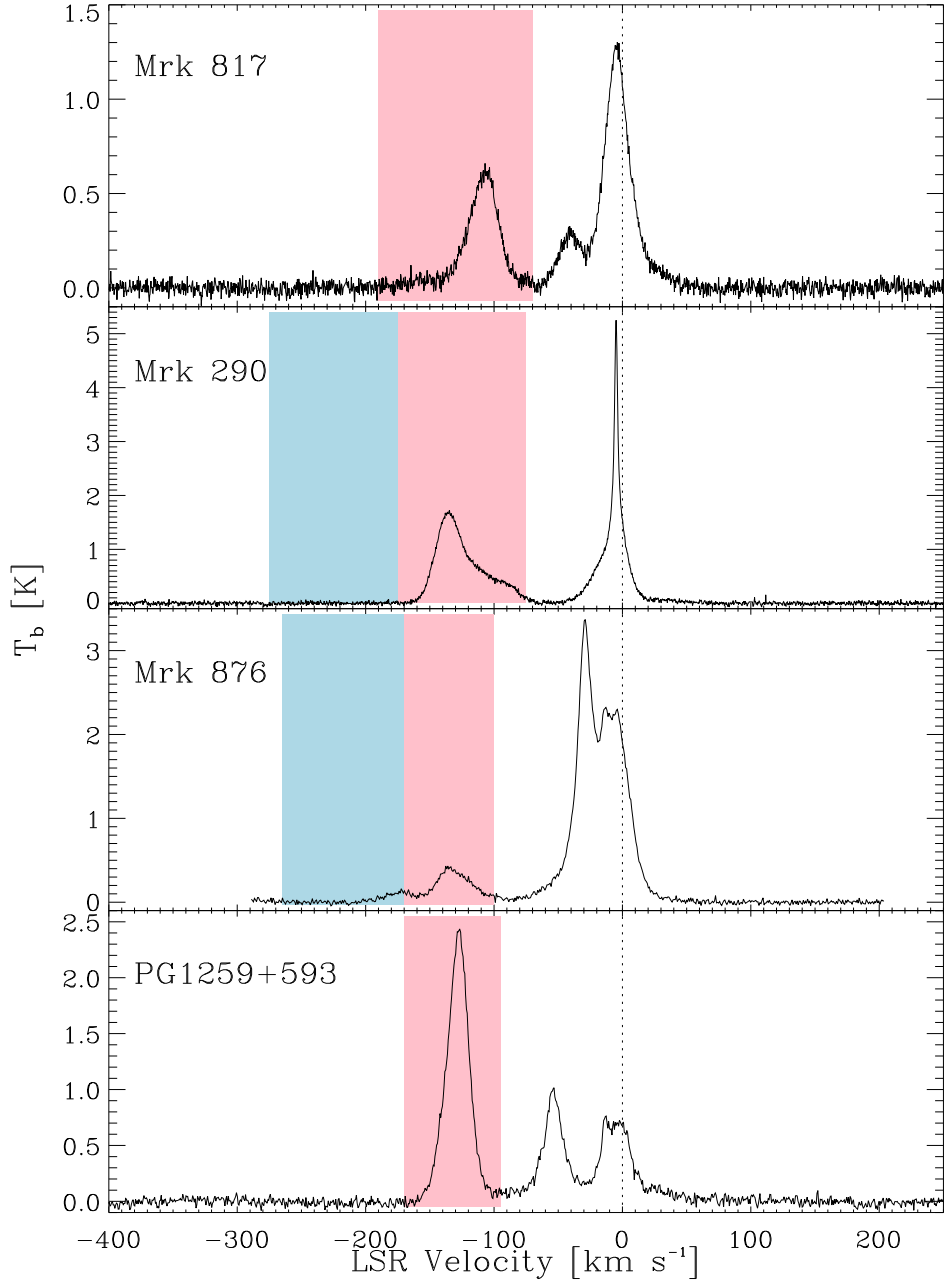}    
   \caption{Spectra of 21-cm emission toward our four target AGN, taken with the GBT.   
     Velocity ranges in pink wash (Complex C) and blue wash (higher velocity gas) 
     show where we detected high-velocity UV absorption with COS:
     Mrk~817 ($-190$ to $-70$~\kms); 
     Mrk~290 ($-175$ to $-75$~\kms\ and $-275$ to $-175$~\kms); 
     Mrk~876 ($-170$ to $-100$~\kms\ and $-265$ to $-170$~\kms);
     PG~1259+593 ($-170$ to $-95$~\kms).  
      }
\end{figure} 

%%%%%%%%%%%%%%%%%%%%%%%%%%%%%%%%%%%%%%%%%%%%

\subsection{Hubble--COS Observations }

Our selected AGN sight lines (Mrk 290, Mrk 817, Mrk~876, PG~1259+593) are the first of
many AGN targets, scheduled over three years of COS guaranteed-time observations that
probe HVCs in Complexes C,  A, M, WD, and WB.   These HVCs were observed previously with 
UV spectrographs aboard the {\it Hubble Space Telescope} (\HST) and the {\it Far Ultraviolet 
Spectroscopic Explorer} (\FUSE), including the \FUSE\ survey of high-velocity \OVI\ (Sembach \etal\ 2003) 
and the \HST/STIS survey of high-velocity \SiIII\ (Shull \etal\ 2009; CSG09).  Previously, 
Mrk 290  was observed with \FUSE\ and \HST/GHRS (Wakker \etal\ 1999; Gibson \etal\ 2001; 
CSG03, CSG07).  
Mrk 817 was observed with \HST/GHRS and \FUSE\  (Gibson \etal\ 2001;  CSG03; Fox \etal\ 2004, 2006).  
PG~1259+593 was observed with \HST/STIS and \FUSE\  (Richter \etal\ 2001; CSG03, CSG07;  
Sembach \etal\  2004); Fox \etal\ 2004, 2006).   
Mrk~876 was observed with HST/GHRS and \FUSE\  (Gibson \etal\ 2001; CSG03, CSG07; Fox \etal\ 2004; 
Sembach \etal\  2003).   

Table~1 lists the relevant COS observational parameters of the current COS program. Our targets were observed in both the G130M (1134--1480~\AA) and G160M (1400--1796~\AA) medium-resolution gratings ($R \approx 18,000$, $\Delta v  \approx 17$ \kms).  The NUV-imaging target acquisitions were performed with the MIRRORA/PSA mode;  the COS primary science aperture ($\sim2.5''$ diameter)
yields good centering  to maximize throughput and resolving power. 
Mrk\,817 was observed during two 2009 epochs (August 4 and December 28).  The first-epoch observations were part of the Early Release Observations (ERO) program (11505, PI: Noll), and the second epoch of observations were part of a COS Guaranteed Time Observations (GTO) program (11524, PI: Green).  The first observations were obtained during the Servicing Mission Orbital Verification (SMOV) period before the instrument reached final focus.  However, the pre-focus data are largely indistinguishable from that taken after correct focus had been achieved except in the case of the very narrowest of absorption features ($\Delta \lambda \la 0.1$~\AA), none of which are analyzed in this work.  Another symptom unique to the early SMOV data was the ``divot and clod'' feature which occurred near the blue end of each detector segment; charge resulting from photons hitting one part of the detector appeared in a different part.  The high voltage on the detectors was reduced on 2009, August 4, and the instrumental feature disappeared.  For the Mrk\,817 ERO observations, the divot and clod were carefully fitted with Gaussian profiles and normalized to the flux level in adjacent portions of the spectrum.   The observing time was approximately equal at both epochs, but the AGN continuum flux decreased by a factor of 1.4 in the interim.  Data from the second epoch were scaled multiplicatively to the flux level of the 2009 August observations.   A full discussion of the Mrk~817 data appears in Winter \etal\ (2011).  Targets  Mrk\,290 and PG\ 1259$+$593 were both observed as part of COS/GTO programs 11524 (2009 October 29) and 11541 (2010 April 15) respectively.  Mrk\,876 was observed (2009 April 8--10) in our COS/GTO program and as a Guest Observer (GO) target (11686, PI: Arav).  The observations were taken within two days of each other.  

After retrieval from the STScI, all data were reduced with the COS calibration pipeline, {\sc CALCOS}\footnote{See the HST Cycle 18 COS Instrument Handbook for more details:\\ {\tt http://www.stsci.edu/hst/cos/documents/handbooks/current/cos\_cover.html}} v2.11f.  The analog nature of the COS micro-channel plate makes it susceptible to temperature changes.  Electronically injected pulses (stims) at opposite corners of the detector allow for the tracking and correction of any drift (wavelength zero-point) and/or stretch (dispersion solution) of the recorded data location as a function of temperature.  
Flat-fielding, alignment, and coaddition of the processed COS exposures were carried out using IDL routines developed by the COS GTO team specifically for COS FUV data\footnote{See {\tt http://casa.colorado.edu/$\sim$danforth/costools.html} for our coaddition and flat-fielding algorithm, which was further described in Danforth et al. (2010).}  The data were corrected for narrow instrumental features arising from shadows from the ion repellor grid wires.  We aligned each exposure by cross-correlating strong ISM features and interpolated the aligned exposures onto a common wavelength scale.  Wavelength shifts were typically on the order of a resolution element ($\sim0.07$~\AA, 17~\kms) or less.  The coadded flux at each wavelength was taken to be the exposure-weighted mean of flux in each exposure.  

To transfer the COS data to the $V_{\rm LSR}$ scale of the 21-cm data, we first aligned the interstellar absorption features in velocity space.  Next we chose between 8 and 12 clearly defined HVC absorption features, found the velocity at which their fluxes were minimized, and took the mean.  This mean velocity was subtracted from the velocity of the \HI\ 21-cm HVC emission peak, and the resulting difference was used to shift the COS data to align with the 21-cm data.  To quantify the quality of the combined data, we identify line-free continuum regions at various wavelengths, smooth the data by the seven-pixel resolution element, and define $(S/N)_{\rm res} =$ mean(flux)/stddev(flux).  
The S/N varies across the wavelength range, but representative values in G130M and G160M data are shown in Table~1.  

Previous observations from \FUSE\ and \HST/GHRS and \HST/STIS (CSG03, CSG07; Fox \etal\ 2004, 2006) were reported on all four sight lines (Mrk\,290, Mrk\,817, Mrk\,876, PG\,1259).  We included \FUSE\ data (HVC lines of \CIII\ and \OVI) and remeasured \OVI\ $\lambda$1031.93, using the velocity range defined by the HVC absorbers measured by COS.  With the higher S/N, these spectra provide better definition of the extent of high-velocity UV absorption.   Tables 3--6 provide the equivalent widths, $W_{\lambda}$, absorption-line data, and derived column densities.  Our error bars include statistical fluctuations in the measurement of equivalent widths and systematic errors arising from variations in the continuum placement and HVC velocity range of integration.   Our measurements of equivalent widths vary the
continuum placement by one standard deviation in continuum flux and adjust the velocity range of HVC absorption by 10~\kms.
These errors are then added in quadrature to produce $1 \sigma$ confidence limits on our AOD measurements.  

The \HST/COS data on \PII\ $\lambda1152.82$ and the \NI\  triplet $\lambda 1199.54,1200.22,1200.70$ (toward Mrk 817, Mrk 876, PG\,1259) are much better determined than the \FUSE\ upper limits.  Other lines in common include \FeII\ $\lambda 1143.22, 1144.94$.   The complex line profiles of \SII\ and \SiII\ are greatly improved in quality, particularly the  \SII\ lines toward Mrk\,876 and Mrk\,290.  We are also able to provide reliable measurements of \NI, \PII, \AlII, and key high ions (\CIV, \SiIV, \NV).

\subsection{GBT Observations}

The 21 cm \HI\ data used here were all obtained with the Robert C.\ Byrd Green Bank Telescope (GBT) of the NRAO at an angular resolution of 9\arcmin.  The spectrum for Mrk~876 is that published by Wakker, Lockman, \& Brown (2011),
the spectrum toward PG\,1259+593 was obtained from archival GBT data, and new observations were made 
for Mrk~817 and Mrk~290.  In all cases, the data were reduced, calibrated and corrected for stray radiation
following the method outlined in Blagrave, Lockman, \& Martin (2010) and Boothroyd et al. (2011, in preparation).  
This technique has been shown to produce 21-cm spectra of HVCs, with errors of a few percent in total \NHI\  limited by noise and residual instrumental baseline effects.  We fitted a second-order polynomial to emission-free regions of each spectrum and smoothed the spectra to produce the following velocity resolution and rms noise:  Mrk~876 (0.81 \kms, 19 mK);  PG\,1259+593 (0.64 \kms, 30 mK);  Mrk~817 (0.32 \kms,  27 mK); and Mrk~290 (0.32 \kms, 25 mK).  

Table 2 presents Gaussian fits to the 21-cm HVC components, including their velocities, widths,  and \HI\ column densities.  
We calculated \HI\ column densities for the HVCs in the optically thin assumption (see footnote to Table 2), which introduces a 
negligible error for the weak HVC lines. The free parameters are the line centers ($V_{\rm LSR}$) of the Gaussians, and their 
height (brightness temperature $T_b$) and full-width at half maximum (FWHM in \kms).   We detected \HI\ emission from all HVCs 
seen in the COS data, except for the most negative velocity component ($V_{\rm LSR} = -275$~\kms\  to $-175$ \kms) toward 
Mrk~290.  For this range, we estimate an upper limit from the noise in the 21 cm spectrum integrated over the velocity range 
of the HVC.

All these directions were observed previously in the 21\,cm line by Wakker \etal\ (2001) using the Effelsberg radio telescope
($9'.7$ beam).  The GBT data ($9'.0$ beam) have lower noise by factors of 2 to 4, slightly better angular resolution, and better 
spectral baselines.  The higher GBT signal-to-noise allows us to fit two components to the HVC toward PG\,1259+593, whereas 
Wakker \etal\ (2001) fitted only one.  
In general, the two telescopes give similar values of \NHI,  to within the errors, except for the $-136$~\kms\ component of Mrk\,290,
where the GBT value is 20\% below that from Effelsberg, and the $-131$~\kms\ component of Mrk 876, where the GBT value
lies 32\% above.  In the latter case, the Effelsberg spectrum appears to have been affected by interference.  As noted in the
captions to Tables 3--6, the total HVC column densities, $\log N_{\rm HI}$,  used in this paper (GBT) are similar to those from 
Effelsberg (Eff):  
Mrk\,817 ($19.50\pm0.01$ from GBT, $19.51\pm0.01$ from Eff);  
Mrk\,290 ($20.05\pm0.01$ from GBT, $20.10\pm0.02$ from Eff);  
Mrk\,876 ($19.39\pm0.02$ from GBT, $19.30\pm0.02$ from Eff);  
PG\,1259+593 ($19.97\pm0.02$ from GBT, $19.95\pm0.01$ from Eff).  
The primary differences are in the emission-line profiles.

%%%%%%%%  Section 3  %%%%%%%%%%%
\section{Scientific Results and Analysis} 

\subsection{Spectra of HVC Absorption Lines}

To illustrate the complexities of identifying an HVC in actual UV absorption-line spectra,
it is helpful to show true fluxes vs.\ wavelength, before deriving equivalent widths from
flux-normalized spectra.   The UV spectra of AGN are rich in interstellar lines of metal ions, but they often 
contain low-redshift IGM absorbers.  The interstellar absorbers include low-velocity gas near the local 
standard of rest (LSR), as well as  high-velocity and intermediate-velocity gas.  

Figures 4--7 present 12-panel plots showing a standard set of HVC absorbers, ranging from 
low to high ionization state.  The velocities of Complex-C absorption are shown in pink wash. 
Along two sight lines (Mrk 290 and Mrk 876), additional HVCs at higher negative velocity are 
shown in blue wash.   We show selected interstellar spectral features for the four AGN sight lines:
absorption lines of eight low-ionization species (\CII, \NI, \OI, \SII, \SiII, \FeII, \AlII, \PII) and 
four more highly ionized species (\SiIII, \SiIV, \CIV, \NV).  We also re-analyzed the \OVI\ data
from \FUSE, using the velocity ranges of HVC ultraviolet absorption, defined by absorption lines 
of \NI, \OI, \SiII, \SII, \FeII, and \AlII.
In many panels, we show multiple transitions of the same species, including the \NI\ triplet 
(1199.550, 1200.223, 1200.710~\AA), the \SII\ triplet (1250.584, 1253.811, 1259.510~\AA), 
the doublets of \CIV\ (1548.195, 1550.770~\AA), \SiIV\ (1393.755, 1402.770~\AA), and \NV\ 
(1238.821, 1242.804~\AA), two \SiII\ lines (1190.416, 1193.290~\AA), and two lines from the 
\CII\ ground state (1334.532~\AA) and \CII$^*$ fine-structure state (1335.663~\AA).  One panel 
shows lines from \OI\ (1302.168~\AA) and \SiII\ (1304.370~\AA), and other panels show lines 
of \SiII\ (1526.707~\AA) and \FeII\ (1143.226, 1144.938~\AA,  and 1608.451~\AA).  

The HVC absorption lines are analyzed to provide absorption equivalent width, $W_{\lambda}$,  
effective line width ($b_{\rm width}$), and column density, $N$.  The column densities were 
derived using the ``apparent optical depth" (AOD) method (Sembach \& Savage 1992), which yields 
the column density, $N_a(v)$, in a given velocity interval.  We then derive the total column density by 
integrating $N_a(v)$ over the velocity range of the HVC.  A few of the lines (\SiIII, \AlII) are mildly saturated, 
but we have checked the column densities from the AOD method with the CoGs for these sight lines,
determined from the low ions (CSG03) for accuracy.  We believe they are reasonable estimates;
further discussion is given in Section 3.4.  

The COS line spread function (LSF) has broad wings with significant power, as documented on the 
STScI/COS website in an article (Kriss \etal\ 2011, COS ISR2011-01) entitled ``Improved Medium 
Resolution Line Spread Function for the COS FUV Spectrum".  Our COS data-reduction software 
accounts for the LSF when fitting Voigt profiles to absorption lines, but the LSF is not included in
the AOD method for deriving column densities, $N_a(v)$, in fixed velocity bins.  Thus, some of the 
increased velocity range could be instrumental in nature, rather than due to the physical nature of the 
gas itself.  However, the differences in velocity and \OVI\ column densities are not large, as
we discuss in the notes on individual sight lines  (Section 3.5).  

Figures 8--11 show ``stack plots" of the HVC absorbers, aligned in velocity space using normalized 
continua.  Our previous experience with the COS wavelength scale (Osterman \etal\ 2011) suggests
that differential shifts up to 10 \kms\ are possible, relative to the 21 cm emission data.  In general, 
the HVC absorption in the UV metal lines agrees well with the \HI\ emission, with the exception of
the Mrk~876 sight line.   In that case, we integrate the UV absorption in Complex C over the range
$V_{\rm LSR} = -170$~\kms\ to $-100$~\kms, whereas the \HI\ emission appears to have a dip
at $\langle V_{\rm LSR} \rangle = -160$~\kms, separating Complex C from higher-velocity absorption
centered at $\langle V_{\rm LSR} \rangle = -190$~\kms\ (see Figures 6 and 10).  

From the normalized COS absorption-line data, we derive equivalent widths, line widths, and  column 
densities, which are displayed in Tables 2--5.  Details on some of the judgements
made in determining column densities are given in Section 3.4.  Tables 7--10 compare previous 
measurements of ion column densities, $\log N(X_i)$, together with our adopted values.  We also show 
the ion ratios referenced to $\log N_{\rm HI}$ and the corresponding elemental abundances.  

\subsection{HVC Column Densities and Abundances}

Tables 7--10 list the adopted and abundances, relative to solar values taken from a recent
review (Asplund \etal\ 2009).  On a logarithmic scale where hydrogen is 12.00, these solar
values are:  C (8.43), N (7.83), O (8.69), S (7.12), Si (7.51), Fe (7.50), Al (6.45), and P (5.41).  
The previous Complex-C abundance estimates were derived from [\OI/\HI] and [\SII/\HI].  
The \OI\ measurements should provide an accurate determination of [O/H], owing to the strong 
resonant charge-exchange coupling of \OI\ and \OII\ with \HII\ and \HI.  However, in many
the sight lines, the \OI\ line ($\lambda1302.16$) is saturated, and we adopted previous values
from weaker lines in the \FUSE\ band.  Abundances from other ions such as [\SII/\HI], [\SiII/\HI], 
and [\FeII/\HI] require ionization corrections to arrive at the true metallicities of these elements.    
There is no guarantee that all elements have the same metallicity, owing to the chemical history 
of the HVC or possible dust depletion of refractory elements (Si, Al, Fe).  As shown in our previous work
(see Fig.\ 16 of CSG03), these corrections are typically expressed as the logarithmic difference 
between the ion abundance and elemental metallicity.  For example, the difference
 [\SII/\HI]$-$[S/H]  ranges from 0.1--0.6 for \HI\ column densities $\log N_{\rm HI} =$ 19.4--20.1
 and HVC physical densities $n_H =$ 0.01--0.1 cm$^{-3}$.  The HVC metallicities are usually
 found from [\OI/\HI], which requires no ionization correction (CSG03, CSG07), and from  [\SII/\HI], 
 reduced by a factor that depends on the \HI\ column density.  
 
 Our four AGN sight lines fall into two distinct groups:  one with low column density
 (Mrk 876 and Mrk 817 have $\log N_{\rm HI} \approx 19.4-19.5$) and one with high
 column density (Mrk 290 and PG\,1259+593 have $\log N_{\rm HI} \approx 20.0-20.1$).  
The ionization corrections to the metal ions (CSG03) depend on the ``photoionization parameter", 
the ratio of ionizing radiation field to gas density $n_H$.  As seen in Figure 1, the HVC clumps 
have characteristic angular sizes of $1^{\circ}-2^{\circ}$, perhaps poorly characterized owing
to the $0.6^{\circ}$ beam and $0.5^{\circ}$ sampling grid of the LAB survey.  
At the 10 kpc distance of Complex~C, one degree corresponds to 175~pc.  If the \HI\ absorbers 
have a comparable depth and angular extent, the observed \HI\ column densities correspond to 
physical densities $n_H \approx 0.05$~cm$^{-3}$ (Mrk 876 and Mrk 817) and  
$n_H \approx 0.2$~cm$^{-3}$ (Mrk 290 and PG\,1259).  From our previous 
photoionization modeling and observed \HI\ column densities, we adopt the following 
corrections.  For Mrk 290 and PG\,1259, the corrections are minor: [\SII/\HI] and [\SiII/\HI] 
are reduced by 0.04 (dex) and [\FeII/\HI] is reduced by 0.02 (dex).  
For Mrk 876, we reduce [\SII/\HI] by 0.32, [\SiII/\HI]  by 0.27, and [\FeII/\HI]  by 0.14. 
For Mrk 817, we reduce [\SII/\HI] by 0.27, [\SiII/\HI]  by 0.21, and [\FeII/\HI]  by 0.11. 
We have not made ionization corrections for \OI\, \NI, \AlII, or \PII.    The ionization 
corrections noted above for \SII, \SiII, and \FeII\ have been applied and included in the 
final column in Tables 7--10.  

\subsection{Inferred Column Densities of Photoionized and Hot Gas}  

The \SiIII\ column density can be used as a proxy for ionized gas to infer the column 
density, N(\HII), of ionized hydrogen that is kinematically associated with the observed \HI.  
As discussed by Shull \etal\ (2009), the \SiIII\ ion likely includes contributions from both
photoionized and collisionally ionized gas (only 16.34 eV needed to produce \SiIII).  We 
adopt an ionization fraction $f_{\rm SiIII} = 0.7 \pm 0.2$ characteristic of multiphase conditions.  
Following the methodology of our \SiIII\ survey with \HST/STIS (Shull \etal\ 2009), we assume a 
metallicity $Z_{\rm Si} \approx 0.1$ relative to the solar abundance, 
(Si/H)$_{\odot} = 3.24 \times 10^{-5}$, to find:
\begin{equation}
   N_{\rm HII} = (4.4 \times 10^5) N_{\rm SiIII}   
              \left[ \frac {Z_{\rm Si}} {0.1}   \right] ^{-1}
              \left[ \frac {f_{\rm SiIII}}  {0.7} \right] ^{-1}   \; .
\end{equation}
For the current four sight lines, we infer ionized column densities, log~N(\HII) 
$= 18.79$ (Mrk\,290), 19.24 (PG\,1259+593), 19.38 (Mrk\,817),  and 19.56 (Mrk\,876).  
Comparing to the observed \HI, these correspond to ionization fractions, N(\HII)/N(\HI),
ranging from low values of 6--10\% in the high-column HVCs toward Mrk\,290 
(log N$_{\rm HI} = 20.05$) and PG\,1259+593 (log N$_{\rm HI} = 19.97$) to much larger 
values of 80\% and 150\% toward Mrk\,817 (log N$_{\rm HI} = 19.50$) and Mrk\,876
(log N$_{\rm HI} = 19.39$).  Interestingly, the highest ionization fractions occur in the
HVCs with the lowest \HI\  column densities.  This suggests that the {\it total} hydrogen
column density distribution is smoother than indicated by the 21-cm emission maps.  

Using arguments similar to the \SiIII\ case, we can use the observed column
densities of high ions (e.g., \CIV\ and \OVI) to infer the column densities of hot, collisionally 
ionized gas.   The COS data on high ions (\NV, \CIV, \OVI, \SiIV, \SiIII) are new or improved over
earlier measurements.  Our re-measurements of \OVI\ HVC absorption are compared 
in Section 3.4 to previous  \FUSE\ studies (Sembach \etal\ 2003; Fox \etal\ 2004; CSG07). 
In many cases, the COS data have higher S/N, which we use to define the velocity range of 
HVC absorption.   We remeasured \OVI\ column densities using these ranges
(see captions to Figures 4--7).  The small differences with previous values (Sembach \etal\ 2003; 
Fox \etal\ 2004) are generally within the stated error bars.  
We adopt metallicities $Z_{\rm C}$ and $Z_{\rm O}$ of 10\% solar  and assume
hot-gas ionization fractions $f_{\rm CIV} = 0.3$ and $f_{\rm OVI} = 0.2$.  The observed \CIV\ 
and \OVI\ column densities toward three sight lines (Mrk 817, Mrk 290, Mrk 876) yield 
consistent  ``hot gas" column densities log~N(\HII) $= 18.7-18.8$, with  \CIV/\OVI\ 
abundance ratios consistent, at N(\CIV)/N(\OVI) = 0.4--0.5.  
The fourth sight line (PG\,1259+593) has the lowest \CIV\ and \OVI\ column densities,
resulting in a lower inferred log~N(\HII) $= 18.2-18.3$ and a slightly lower ratio,
N(\CIV)/N(\OVI) = 0.3.  

The ionization ratios, \CIV/\OVI,  \NV/\OVI, \SiIV/\OVI, can be compared with observations
of highly ionized HVCs (Fox \etal\ 2004, 2005, 2006;  CSG05; CSG07; Indebetouw \& Shull 2004b) 
and with models of various ionization processes (see Figure 1 of Indebetouw \& Shull 2004a).  
In ratio plots of \CIV/\OVI\ and \NV/\OVI, the regions occupied by HVCs along our four sight lines 
are consistent with hot gas in conductive or radiatively cooling interfaces.  The observed 
ratios are {\it not} consistent with collisional ionization equilibrium (CIE) or turbulent mixing
layers (TML).  For relative solar abundances (O: C: N = 1.00: 0.55: 0.14), the CIE models produce 
much higher \NV/\CIV\  ratios than observed;  the obvious explanation is the lower 
nitrogen abundance observed in neutral HVC gas.  Models of TMLs exhibit a wide range of 
predictions, owing to the assumptions and parameterizations that go into calculating the
ionization and cooling.  Early TML models (Slavin, Shull, \& Begelman 1993; Indebetouw \& Shull 2004) 
assumed mixing to an intermediate temperature, $\overline T \approx 10^5$~K and relaxed to 
ionization equilibrium (Slavin, Shull, \& Begelman 1993;  Indebetouw \& Shull 2004a).
Those models produce much higher \CIV/\OVI\ ratios than observed.  
More recent TML models (Kwak \& Shelton 2010) that incorporate non-equilibrium ionization
find regions with warm, radiatively cooled \CIV, mixed with hotter gas, out of ionization 
equilibrium.  These models find 2--4 times higher column densities in \NV, \CIV, and  \OVI\ abundances
than predicted in CIE.  However, their typical TML ratios are:  
 \CIV/\OVI\ = 1.5 (range 0.8--2.4) and \NV/\OVI\ (range 0.14--0.32), both higher than the 
COS-observations, which have more \OVI\ than predicted by the models.  
Models of non-equilibrium ionization (Gnat \& Sternberg 2007) with time-dependent 
cooling ($Z \approx 0.1 Z_{\odot}$) find fair agreement with the COS-observed  ionization ratios.  

We observe a range of ratios, \SiIV/\CIV\ $\approx$ 0.15--0.36, along the four sight lines, typical of the 
previous studies (Fox \etal\ 2004, 2005;  CSG05; CSG07).   For relative solar abundances, 
(Si/C)$_{\odot} = 0.11$, photoionization models of HVCs  findabundance ratios, 
N(\SiIV)/N(\CIV) $> 1$, for values of photoionization parameter, $\log U = -3.0\pm0.2$, needed to explain
the low ions and fit the ratios \SiIV/\SiIII/\SiII\ (Shull \etal\ 2009).  
These same models also under-predict the total column densities of high ions (\CIV, \SiIV, \NV, \OVI).  
Similarly, photoionization models with the observed range, N(\SiIV)/N(\CIV) = 0.1--0.3, predict 
N(\SiIV)/N(\SiII) $\approx 1$, much higher than observed.  Therefore, it is likely that some \CIV\ and \SiIV\ comes 
from hot gas.  The kinematic association of low and high ions in these HVCs 
requires a mixture of  denser cloud cores of \HI\ with extended warm photoionized gas and
sheaths of much hotter gas, perhaps produced by bow shocks and turbulent or conductive 
interfaces between the HVC core and hot halo gas.   However, detailed comparisons of observations
and models are often complicated by the assumptions of {\it relative} solar abundances.  

\subsection{Toy Model for HVC Clouds in the Galactic Halo} 

One of the reasons for renewed interest in Galactic HVCs is the possibility that the Local Group 
might contain considerable mass in virialized halos and a hot circumgalactic medium.   
Spitzer (1956) first suggested the existence of low-density coronal gas with $T = 10^6$~K and 
$n_e = 5 \times 10^{-4}$~cm$^{-3}$, extending 8 kpc
above the Galactic plane and providing pressure confinement of observed high latitude clouds.  
Kahn \& Woltjer (1959) noted inconsistencies in galactic stellar masses and  Local Group
dynamics (before the inference of dark matter) and suggested the existence of a substantial 
reservoir of low-density halo gas, with $T = 5\times 10^6$~K  and $n_e = 1 \times 10^{-4}$~cm$^{-3}$.  
Direct probes of hot, low-density gas are difficult, owing to the $n^2$ dependence of its X-ray
emission and the contamination of most signals by foreground electrons in the kpc-scale  ``Reynolds 
layer"  (Reynolds 1991).  Indirect probes of the halo density yield limits ($n_e < 10^{-4}$~cm$^{-3}$) 
from the effects of drag on orbits of the Magellanic Stream (Moore \& Davis 1994).    Similar
limits ($n_e < 3 \times 10^{-5}$~cm$^{-3}$) follow from ram-pressure stripping of Local Group 
dwarf galaxies (Blitz \& Robishaw 2000).   More recently, Heitsch \& Putman (2009) used numerical 
simulations to suggest that infalling HVCs with \HI\ masses less than $10^{4.5}\,M_{\odot}$ may
become fully ionized by Kelvin-Helmholtz instabilities within $10^8$~yr ($\sim10$~kpc for
typical HVC velocities of 100 \kms).   All of these estimates depend critically on the assumed halo 
gas density and on dynamical interactions at the boundaries between the HVCs and the hot, low-density 
medium that confines them.

As a toy model for HVC cloud confinement, we consider spherical clouds with radius $R$, mass $M$, 
constant gas density $\rho$, and mean atomic mass $\mu = \rho/n_H = 1.23 m_H$ (helium 25\% by 
mass) in virial equilibrium confined by external pressure, $P_0$.    For clumps of angular radius 
$\theta_{\rm deg}$ (in degrees),  we adopt $R = (175~{\rm pc}) \theta_{\rm deg}$ at 10 kpc distance,
$n_H = (0.1~{\rm cm}^{-3}) n_{0.1}$, and temperature $T = (100~K)T_{100}$. The virial theorem  
with confinement (Spitzer 1978) requires that
\begin{equation}
   4 \pi R^3 \, P_0 = \frac {3MkT}{\mu}  - \frac {3GM^2}{5R}  \; .  
\end{equation}
The confining pressure could arise from hot Galactic halo gas, which we scale to nominal values
$n_{\rm halo} = (10^{-5}$~cm$^{-3})\,n_{-5}$ and $T_{\rm halo}= (10^6~{\rm K})\,T_6$.  Alternately,
the HVCs could be confined by ram pressure, $P_{\rm ram} = \rho_{\rm halo} V_{\rm HVC}^2$,
as they fall through the halo.  For the assumed halo parameters, the total thermal pressure of fully 
ionized gas with $n_{\rm He}/n_{\rm H} = 0.0823$ is 
$P/k  \approx 2.25 n_H T  = (22.5~{\rm cm}^{-3}~{\rm K}) n_{-5}\, T_6$.  This pressure
is consistent with inferences from various highly ionized HVCs (Sembach \etal\ 1999; CSG05),
although such estimates are uncertain owing to assumptions in the ionization modeling.  The
halo density probably lies in the range $n_e = (1-10) \times 10^{-5}$ cm$^{-3}$, with considerable
variation over vertical distances 5--50~kpc above the Galactic plane.  

In our model, the HVC clump masses in Complex C are 
$M = (6.8 \times10^4~M_{\odot}) n_{0.1} \theta_{\rm deg}^3$, and the terms in the virial equation are 
all of comparable size,  
\begin{eqnarray}
    (3MkT / \mu)   &=& (2.7 \times 10^{48}~{\rm erg})  \, n_{0.1}\, T_{100} \, \theta_{\rm deg}^3   \\
    (3GM^2 / 5R) &=& (1.4 \times 10^{48}~{\rm erg})\,  n_{0.1}^2 \, \theta_{\rm deg}^5  \\
   4 \pi R^3 \, P_0      &=&  (4.0 \times 10^{48}~{\rm erg}) \,  [n_{\rm halo}/10^{-5}~{\rm cm}^{-3}] \, \theta_{\rm deg}^3  
       \;  .  
\end{eqnarray}
The thermal pressure of the halo, $P_{\rm halo} \approx 1.6\times10^{-15}$ erg~cm$^{-3}$,  is comparable 
to the ram pressure on the HVCs.  This not surprising, as these infalling clouds have low Mach numbers with 
respect to the hot gas.   The HVC clumps may not be self-gravitating (no \HH\ or stars have been detected).
However, their observed properties place them near virial equilibrium (with $T \leq 10^3$~K), and there may 
be no need to invoke dark matter for their confinement.  

As the HVCs encounter the higher densities in the stratified lower halo, their outer portions may be torn 
apart and dissipated by interface instabilities.  This would feed the Galactic halo rather than replenishing 
the reservoir of star formation in the disk.  Portions of Complex C do appear to be clumping up, although
the individual clump masses ($\theta_{\rm deg} \geq 1$) are likely above the $10^{4.5}\,M_{\odot}$ threshold 
for survival (Heitsch \& Putman 2009).

\subsection{Notes on Individual Sight Lines}  

In this section, we compare the COS  equivalent widths of several lines,  previously
measured by GHRS and STIS spectrographs (CSG03, CSG07; Richter \etal\ 2001). 
The derived column densities of various ions are compared in Tables 7--10.   For some of the
high ions (\CIV, \SiIV, \SiIII, \NV) our COS measurements are the only published data.  
Previous measurements of  high-velocity \SiIII\ toward Mrk~876 and PG\,1259+593
were discussed in the \HST\ surveys by Shull \etal\ (2009) and Collins \etal\ (2009).

\noindent
{\bf (1) Mrk 817.}   The three unsaturated \SII\ lines at 1250, 1253, and 1259 \AA\ should have 
equivalent widths in the ratio 1:2:3 based on their relative oscillator strengths. The weakest line 
($\lambda1250.6$) was measured at $W_{\lambda} = 27\pm8$ m\AA\  (COS) vs.\ $16\pm3$ m\AA\ 
(GHRS).  The stronger line ($\lambda1253.8$) was $19\pm5$ m\AA\ (COS) vs.\ $28\pm3$ m\AA\
(GHRS).   The strongest line ($\lambda1259.5$) with $42\pm7$~m\AA\  (COS) was not reported
with GHRS.  We base our  \SII\ column density, $\log N = 14.29\pm0.08$, on the $\lambda1259$ 
measurement. For other lines we used \SiIV\ $\lambda 1394$, \NV\  $\lambda1238$,  
\SiII\ $\lambda1304$ (consistent with $\lambda1526$), and  \FeII\ $\lambda1608$ (consistent 
with $\lambda1145$).   We find good agreement between both lines in the doublets of \CIV\  
($\lambda\lambda 1548,1551$), \NV\ ($\lambda\lambda 1238,1242$), and \SiIV\ 
($\lambda\lambda1394, 1402$).  Fox \etal\ (2004) suggested that there may be no detectable
high-velocity \NV\  toward Mrk~817 because of  intergalactic \Lya\ absorbers at $z = 0.0189$ 
and $z = 0.0194$.  However, we believe to have detected \NV\ with log~N = $13.16^{+0.12}_{-0.15}$
at the $-109$~\kms\ velocity of Complex~C.  As seen in Figure 8,  the \Lya\ absorber at $z = 0.0184$ 
would appear at $-42$~\kms\ in the \NV\ rest frame, fairly well separated from the HVC.  The ratios
of \NV\ to other high ions (\CIV, \SiIV, \OVI) are typical of other sight lines.  
We use the weaker \NI\ $\lambda1199.5$ line, since the other lines (1200.2~\AA\ 
and 1200.7~\AA) are blended with $V_{\rm LSR} = 0$ absorption.  
Both \AlII\ $\lambda1670.78$ and \SiIII\ $\lambda1206.50$ are mildly saturated, at equivalent
widths of $\sim400$~m\AA\ and $\sim350$~m\AA, respectively.  Our inferred column densities from 
AOD integration of $N_a(v)$ accurately reflect this saturation, for line broadening of 20--30 \kms, 
somewhat higher than the doppler parameter $b = 11$ \kms\  inferred from \OI\ and
other low ions (CGS03).  
Re-measuring the  \OVI\ column density, we find  $\log N = 14.05^{+0.16}_{-0.09}$ compared to 
$13.97^{+0.10}_{-0.13}$ from Fox \etal\ (2004) and $13.88\pm0.20$ (Sembach \etal\ 2003).
The small differences in \OVI\ column densities arise from the velocity range adopted
for the HVC absorption.  From the extent of  UV absorption (\SII, \SiII, \FeII, \AlII) in Figure 8,
we integrate between $V_{\rm LSR} = -190$ and $-70$~\kms,  whereas Fox \etal\ (2004) used the 
interval from $-160$ to $-80$~\kms.

\noindent
{\bf (2) Mrk 290.}   The \SII\ data from COS are superior to that from GHRS or STIS (CSG03, CSG07).
The HVC components of $\lambda1250.57$ and $\lambda1253.80$ lie in the blue 
and red wings, respectively, of the broad \Lya\ emission line of the AGN (see Fig.\ 5).   To arrive 
at the column density, $\log N = 14.43\pm0.08$, we use the $\lambda1259.51$ line. For \SiII, both lines
($\lambda1304$ and $\lambda1526$) are saturated, yielding a lower limit on column density.
We adopt the CoG value, $\log N = 14.9\pm0.15$, from CSG07.  
The column density from the mildly saturated \FeII\  $\lambda1608$ is consistent with $\lambda1145$.  
We see agreement between both lines in the doublets of \CIV\  ($\lambda\lambda 1548,1551$),
\NV\ ($\lambda\lambda 1238,1242$), and \SiIV\ ($\lambda\lambda 1394,1402$).  
We use the weaker line \NI\ $\lambda1199.5$ line, since the other lines (1200.2~\AA\ and 1200.7~\AA)
are blended with $V_{\rm LSR} = 0$ absorption.  
Both \AlII\ $\lambda1670.78$ and \SiIII\ $\lambda1206.50$ are mildly saturated, at equivalent
widths of $\sim340$~m\AA\ and $\sim345$~m\AA, respectively.  Our inferred column densities from 
AOD integration of $N_a(v)$ accurately reflect this saturation, for line broadening of 20--30 \kms, 
somewhat higher than the doppler parameter $b = 16$ \kms\  inferred from \OI\ and
low ions (CGS07).  We re-measured the \OVI\ column density, finding reasonable agreement,
$\log N = 14.10^{+0.15}_{-0.14}$ (between $-175$ and $-75$~\kms) compared to $14.23\pm0.04$ 
(CSG07 integrated between $-165$ and $-75$~\kms) and $14.20\pm0.16$ (Sembach \etal\ 2003).

\noindent
{\bf (3) Mrk 876.}   The \SII\ data from COS are superior to that from STIS (CSG07),
which only gave an upper limit, $\log N < 14.34$.  With COS, the HVC components of 
$\lambda1259.51$ and $\lambda1253.80$ yield essentially the same column density, 
$\log N = 14.25^{+0.13}_{-0.14}$.   For \SiII, both lines ($\lambda1304$ and $\lambda1526$) are 
mildly saturated, but yield a consistent column density, $\log N = 14.41^{+0.16}_{-0.09}$.  
The column density from \FeII\  $\lambda1608$, $\log N = 14.26^{+0.17}_{-0.10}$, is consistent 
with that from $\lambda1145$.  We find agreement between both lines in the doublets of \CIV\  
($\lambda\lambda 1548,1551$), \NV\ ($\lambda\lambda1238,1242$), and \SiIV\ ($\lambda\lambda 
1394,1402$).  We use the weaker \NI\  $\lambda1199.5$ line, since the other lines (1200.2~\AA\ 
and 1200.7~\AA\ are blended with $V_{\rm LSR} = 0$ absorption.  
Both \AlII\ $\lambda1670.78$ and \SiIII\ $\lambda1206.50$ are mildly saturated, at equivalent
widths of $\sim280$~m\AA\ and $\sim270$~m\AA, respectively.  Our inferred column densities from 
AOD integration of $N_a(v)$ accurately reflect this saturation, for line broadening of 20 \kms, 
somewhat higher than the doppler parameter $b = 16$ \kms\  inferred from \OI\ and low ions (CGS03).  
Our previous HVC survey of \SiIII\ (Shull \etal\  2009) measured $\log N \geq 13.92$ for this absorber, 
slightly higher than the value, $13.72^{+0.24}_{-0.14}$ measured here.
We re-measured the  \OVI\ column density, finding moderate differences, but within stated errors,
$\log N = 13.99^{+0.36}_{-0.14}$ compared to $14.20\pm0.02$ (CSG07), $14.05\pm0.17$ 
(Sembach \etal\ 2003), and $14.12^{+0.11}_{-0.13}$ (Fox \etal\ 2004).   From the observed extent
of the UV absorption (\SII, \SiII, \FeII), we integrate \OVI\ between $V_{\rm LSR} = -170$ and 
$-100$~\kms, whereas Fox \etal\ (2004) used the interval from $-220$ to $-100$~\kms, and 
CSG07 used $-210$ to $-75$~\kms.

\noindent
{\bf (4) PG~1259+593.}   The \SII\ data from COS are superior to that from GHRS (CSG03)
and agree well with those from STIS (Sembach \etal\ 2004; Richter \etal\ 2001).  Our column density,
$\log N = 14.35^{+0.15}_{-0.10}$ agrees with previous measurements.   For \SiII, both lines 
($\lambda1304$ and $\lambda1526$) are mildly saturated and yield a consistent column density.
The column density from \FeII\  $\lambda1608$ is consistent with $\lambda1145$.  
We find agreement between both lines in the doublets of \CIV\  ($\lambda 1548,1551$),
\NV\ ($\lambda1238,1242$), and \SiIV\ ($\lambda 1394,1402$).  Neither line in the \NV\ doublet 
($\lambda1238,1242$) is seen, to a limit $\log N < 12.99$ ($3\sigma)$.
We use the weaker \NI\  $\lambda1199.5$ line, since the other lines at 1200.2~\AA\ and
1200.7~\AA\ are blended with $V_{\rm LSR} = 0$ absorption.  
Both \AlII\ $\lambda1670.78$ and \SiIII\ $\lambda1206.50$ are mildly saturated, at equivalent
widths of $\sim215$~m\AA\ and $\sim190$~m\AA, respectively.  Our inferred column densities 
from AOD integration of $N_a(v)$ accurately reflect this saturation, for line broadening of 20--30 \kms, 
somewhat higher than the doppler parameter $b = 10$ \kms\  inferred from \OI\ and low ions (CGS03).  
Our previous HVC survey of \SiIII\ (Shull \etal\  2009) measured $\log N \geq 13.60$ for this absorber, 
somewhat higher than the value, $\log N = 13.30^{+0.24}_{-0.16}$ measured here.
We re-measured the \OVI\ column density of this HVC, finding fair agreement,
$\log N = 13.58^{+0.18}_{-0.12}$ compared to $13.71 \pm 0.09$ (Fox \etal\ 2004) and 
$13.72 \pm 0.17$ (Sembach \etal\ 2003).  We integrate between $V_{\rm LSR} = -170$ and $-95$~\kms, 
while Fox \etal\ (2004) used the interval from $-160$ to $-80$~\kms.  
For the \OI\ column density in Complex C, we adopt the careful measurement, $\log N_{\rm OI} = 15.85 \pm 0.15$
from Sembach \etal\ (2004), who fitted components as part of their study to find a deuterium ratio,
D/H = $(2.2 \pm 0.7)\times10^{-5}$ in this HVC.  

\newpage

%%%%%%  Section 4   %%%%%%%%%%%%
   
\section{Conclusions and Implications} 

The high throughput and low background of the Cosmic Origins Spectrograph allow us to 
observe Complex HVCs with high S/N and improved velocity accuracy.  A summary of the 
major issues and new observational results follows:
\begin{itemize}

\item Comparison of COS/G130M and G160M data with previous HST and FUSE data 
shows a more complete array of transitions and elements, with higher S/N and a better
defined velocity range of UV absorption.  

\item In general, the COS-derived column densities and abundances agree with previous 
UV spectroscopic studies by GHRS, STIS, and FUSE.   The Complex C metallicity inferred 
from \OI\ and \SII\ lies between 10--30\% of solar values.   The better-defined velocity range 
of  UV absorption affects some of the high-ion column densities such as \OVI.  

\item Using \SiIII\ as a proxy for \HII, we find  a substantial amount of ionized gas kinematically 
associated with the \HI\ 21-cm emission, ranging from N(\HII)/N(\HI) $\approx 0.1$ (Mrk\,290,
PG\,1259+593) to 0.8--1.5 (Mrk\,817, Mrk\,876).  The  HVCs are expected to have extended 
ionized atmospheres owing to photoionization, and possibly from interactions with the halo.  

\item Individual clumps of Complex~C appear to be near virial equilibrium with pressure
confinement.  Their masses ($\geq10^5\,M_{\odot}$) are probably above the threshold for
survival against dissipation by interface instabilities. 

\item High ionization states (\CIV, \NV, \SiIV, \OVI) are seen in all four sight lines, with column densities 
ranging over factors of 3--5 and ratios consistent with multi-phase ionization processes:
N(\SiIV)/N(\OVI) $\approx$ 0.05--0.11, N(\CIV)/N(\OVI) $\approx$ 0.3--0.5, 
N(\NV)/N(\OVI) $\approx$ 0.07--0.13, and N(\SiIV)/N(\SiIII) $\approx$ 0.1--0.2.  The observed ratios
\SiIV/\CIV\ $\approx$ 0.15--0.36 are inconsistent with photoionization models for the low ions.  

\item Models of high-ion production and \CIV, \NV, \OVI\ line ratios in Galactic halo gas 
(Indebetouw \& Shull 2004a,b; Gnat \& Sternberg 2007) suggest radiative cooling and 
conductive heating.  They are inconsistent with collisional ionization equilibrium, which produces
more \NV\ and less \OVI\ than observed:  N(\NV)/N(\OVI) $\approx 0.3$ and 
N(\CIV)/N(\OVI) $\approx 0.1$.  The low \NV\ probably reflects the nitrogen underabundance
observed in the neutral components of many HVCs.  

\end{itemize}

As noted in the Introduction, Complex C is a large ($M \sim 10^7~M_{\odot}$) HVC with substantial 
amounts of neutral and ionized gas, falling toward the Galactic plane with an average mass flow of 
$0.1 M_{\odot}~{\rm yr}^{-1}$ of low-metallicity gas (10--30\% solar).  
Over the entire Galactic halo, as probed by \SiIII\ (Shull \etal\ 2009), HVCs may provide sufficient 
gaseous infall, $\sim 1~M_{\odot}~{\rm yr}^{-1}$ to help replenish some of the 2--4~$M_{\odot}~{\rm yr}^{-1}$ 
of star formation in the disk.  Future HVC studies with COS will include sight lines passing through 
Complex M (Danly \etal\ 1993; Yao \etal\ 2011), Complex A (van Woerden \etal\ 1999), and the Smith 
Cloud (Lockman \etal\ 2008).  With more accurate metal-ion abundances and ionization corrections, 
we may soon be able to assess the relative abundance ratios of different elements (S, Si, O, Fe, N, 
and perhaps P and Al)  seeking to find non-solar ratios indicative of chemical history.  
Nitrogen is underabundant relative to C and S, which suggests that Complex C gas is chemically young.

The abundances of refractory elements (Si, Al, Fe) relative to undepleted S could limit the amount of 
dust, which is expected to be small, given the lack of clear evidence for infrared emission from
Complex C.  Typical values for these four sight lines are (S/Fe) $\approx$ 1.0--1.7, (Si/Fe) $\approx$ 1.2--1.7, 
and (Al/Fe) $\approx$ 0.06--0.10.  Because this range is within the uncertainty of the measurements,
it is difficult to draw any firm conclusions.  
The [$\alpha$/Fe] ratios are interpreted as the ratio of massive-star nucleosynthesis (and 
core-collapse SNe) to nucleosynthesis from older stars (and Type Ia SNe, which produce more Fe).

%%%%%%%%%%%%%%%%%%%%%%%%%%%%%%%%%%%%%%%%%%%%%%%%%%%%%%%%%  

\smallskip

%%%%%%%%%%%%%%%%%%%%%%%%%%%%%%%%%%%%%%%%%%%%%%%%%%%%%%%%%%

\acknowledgments

It is our pleasure to acknowledge the thousands of people who made HST Servicing Mission 4 
a huge success.  We thank Brian Keeney,  St\'ephane B\'eland, and the rest of the COS/GTO team 
for their work on the calibration and verification of the early COS data.   We thank Andrew Fox for
helpful discussions of his previous studies of the high ions toward Complex~C and Mark Giroux
for insights on photoionization modeling of HVCs.    We also appreciate comments from the referee, 
Kenneth Sembach, on physical conditions in HVCs and details of the COS data analysis.  
This work was supported by NASA grants NNX08AC146 and NAS5-98043 
and the Astrophysical Theory Program (NNX07-AG77G  from NASA and
AST07-07474 from NSF) at the University of Colorado at Boulder.

%%%%%  Table 1 (Observing Log)  %%%%%%%%%
%%% !!! add S/N
%%% !!! add coords
\begin{deluxetable}{lcccccc}
\tabletypesize{\footnotesize}
\tablecaption{COS Observations}
\tablecolumns{7}
\tablewidth{0pt}
\tablehead{
  \colhead{Target\tablenotemark{a}} &
  \colhead{Coordinates\tablenotemark{a}} &
  \colhead{Program(s)}  &
  \colhead{Grating}   &               
  \colhead{$N_{\rm exp}$ }    & 
  \colhead{$t_{\rm exp}$ (s)}  &  
  \colhead{$(S/N)_{\rm res}$\tablenotemark{b}}  
            }
 \startdata 
 Mrk\,817                   & 14:36:22.1 $+$58:47:39  & 11505, 11524  & G130M     & 8 &  3426   & 41 \\
$z = 0.031455$       & $l=100.30,~b=53.48$      & 11505, 11524  & G160M     & 8 &  3010   & 35 \\
 Mrk\,290                   & 15:35:52.3 $+$57:54:09  & 11524                & G130M     & 4 &  3857  & 22 \\
$z = 0.029577$       & $l=~91.49,~b=47.95$      & 11524                & G160M     & 4 &  4801  & 22 \\
Mrk\,876                   & 16:13:57.2 $+$65:43:10  & 11524, 11686  & G130M     & 6 & 12580 & 57 \\
$z = 0.12900$         & $l=~98.27,~b=40.38$       & 11686                & G160M     & 4 & 11820 & 37 \\
PG\,1259$+$593    & 13:01:13.1 $+$59:02:06  & 11541                &  G130M    & 4 &  9201  & 35 \\
$z = 0.4778$            & $l=120.56,~b=58.05$      & 11541                & G160M     & 4 & 11169 & 32 \\
\enddata

\tablenotetext{a}{AGN target, redshifts and coordinates: RA/Decl (2000) and Galactic $\ell$ and $b$. }

 \tablenotetext{b}{Signal-to-noise ratio per resolution element calculated at 1350 \AA\ and 
 1550 \AA\ for data with G130M (1132--1460~\AA) and G160M (1394--1798~\AA) gratings,
 respectively.  The actual S/N values vary throughout each dataset, but the quoted values 
 give a sense of relative data quality.  See text for more details.}
\end{deluxetable}

%%%%%%%%%%%%%%%%%%%%%%%

%%%%%  Table 2 (Gaussian Fits to 21-cm Data)  %%%%%%%%%

\begin{deluxetable}{lllll}
\tabletypesize{\footnotesize}
\tablecaption{Gaussian Fits\tablenotemark{a} to \HI\ Column Densities}
\tablecolumns{5}
\tablewidth{0pt}
\tablehead{
      \colhead{Target}       &  \colhead{$\langle V_{\rm LSR} \rangle$}    &
       \colhead{Height}      &  \colhead{FWHM}    &    \colhead{$N_{\rm HI}$}    \\
\colhead{}     &   \colhead{(\kms)}  &   \colhead{($T_b$(K))}  &
   \colhead{(\kms)}  &   \colhead{(in $10^{18}$~cm$^{-2}$)}    
       }
\startdata 
Mrk\,817                    & $-108.0\pm0.12$       &  $0.594\pm0.0055$    &  $27.5\pm0.3$  & $31.7\pm 0.5$   \\
Mrk\,290                    &  $-89.8\pm1.2$           &   $0.310\pm0.030$     &  $22.4\pm1.3$   &  $13.5\pm 1.5$   \\                          
                                    & $-113.6\pm0.5$         &   $0.583\pm0.021$     &  $25.7\pm2.8$   &  $29.1\pm 3.3$   \\
                                    & $-136.3\pm0.26$       &  $1.608\pm0.037$      &  $22.2\pm0.3$  &  $69.4\pm 1.8$   \\
Mrk\,876                    & $-131.4\pm0.3$         &   $0.381\pm0.006$     &  $33.4\pm0.7$   &  $24.7\pm 0.7$   \\
                                    & $-175.5\pm0.8$         &   $0.116\pm0.008$     &  $22.7\pm1.9$   &  $  5.1\pm 0.5$   \\
PG\,1259$+$593     & $-126.4\pm0.1$         &   $ 0.891\pm0.061$    &  $12.0\pm0.5$   &  $20.7 \pm 1.6$      \\
                                    & $-129.1\pm0.1$         &   $1.558\pm0.062$ &  $24.1\pm0.3$   &  $73.0\pm 3.1$      \\

\enddata

\tablenotetext{a}{Gaussian components fitted to three parameters:  LSR velocity ($V_{\rm LSR}$),  
  height ($T_b$), and full-width half-maximum (FWHM), from which we derive column density,
  $N_{\rm HI} = (1.823 \times 10^{18}$ cm$^{-2}$)(height)(FWHM)(1.065), for height in units of 
  $T_b$ (K) and  FWHM in \kms.  
    }
\end{deluxetable}

%%%%%%%%%%%%%%%%%%%%%%%

%%%%%
%%%%%  Table 3 -- revised by Matt Stevans  (HVC data for Mrk 817)  %%%%%%%%%
%%%!!! sig figs
\begin{deluxetable}{llllll}
\tabletypesize{\footnotesize}
\tablecaption{HVC Measurements\tablenotemark{a} (Mrk 817) }
\tablecolumns{6}
\tablewidth{0pt}
\tablehead{
   \colhead{Ion}   &
   \colhead{$\lambda_{0}$}   &
   \colhead{$V_c$} &
   \colhead{$b_{\rm width}$}   &
   \colhead{$W_{\lambda}$} &
   \colhead{$\log N_a$}    \\
   \colhead{}        &
   \colhead{(\AA)}   & 
   \colhead{(\kms)}  &
   \colhead{(\kms)}  &
   \colhead{(m\AA)}  &
   \colhead{($N$ in cm$^{-2}$) }    
 }

\startdata	
  S\,II   &   1259.51   &$-104\pm11$&$   25\pm 7  $&$    42\pm 7  $&$   14.29^{+0.08}_{-0.08}$ \\
  S\,II   &   1253.80   &$-102\pm 7$&$   27\pm 9    $&$    19\pm 5  $&$   14.12^{+0.12}_{-0.12}$ \\
  S\,II   &   1250.57   &$-123\pm 1$&$   49\pm14   $&$    27\pm 8  $&$   14.57^{+0.15}_{-0.13}$ \\
  Si\,II  &   1526.70   &$-121\pm 5$&$   43\pm 9    $&$   364\pm66  $&$   14.37^{+0.08}_{-0.07}$ \\
  Si\,II  &   1304.37   &$-114\pm 6$&$   38\pm 6    $&$   241\pm39  $&$   14.47^{+0.08}_{-0.05}$ \\
  Si\,II  &   1193.28   &$-116\pm 6$&$   41\pm 7    $&$   338\pm61  $&$   14.08^{+0.09}_{-0.07}$ \\
  Si\,II  &   1190.41   &$-116\pm 5$&$   40\pm 8    $&$   295\pm57  $&$   14.23^{+0.10}_{-0.07}$ \\
  Fe\,II  &   1608.45   &$-113\pm 7$&$   38\pm 6   $&$   173\pm32  $&$   14.23^{+0.10}_{-0.06}$ \\
  Fe\,II  &   1144.93   &$-113\pm 6$&$   40\pm 7   $&$   117\pm25  $&$   14.19^{+0.11}_{-0.09}$ \\
  Fe\,II  &   1143.22   &$-117\pm 2$&$   50\pm17  $&$    49\pm14  $&$   14.40^{+0.17}_{-0.12}$ \\
  C\,II*  &   1335.66   &$-141\pm 5$&$   55\pm16  $&$    31\pm15  $&$   14.22^{+0.26}_{-0.19}$ \\
  C\,II   &   1334.53   &$-121\pm 5$&$   43\pm 7    $&$   431\pm75  $&$   14.90^{+0.12}_{-0.12}$ \\
   N\,I    &   1200.71   &$-131\pm 1$&$   42\pm 9   $&$   284\pm44  $&$   14.95^{+0.05}_{-0.04}$ \\
  N\,I    &   1200.22   &$-138\pm 2$&$   47\pm11    $&$   284\pm68  $&$   14.65^{+0.12}_{-0.08}$ \\
  N\,I    &   1199.54   &$-109\pm 5$&$   36\pm 6    $&$   141\pm27  $&$   14.06^{+0.10}_{-0.07}$ \\
  O\,I    &   1302.16   &$-119\pm 5$&$   42\pm 7    $&$   337\pm57  $&$   14.98^{+0.07}_{-0.06}$ \\
  Al\,II  &   1670.78   &$-119\pm 6$&$   40\pm 7    $&$   404\pm65  $&$   13.23^{+0.07}_{-0.06}$ \\
  P\,II   &   1152.82   &$-117\pm 8$&$   37\pm 9    $&$    23\pm 6     $&$   12.95^{+0.09}_{-0.17}$ \\  
  C\,IV   &   1550.78   &$-107\pm 9$&$   37\pm 4  $&$   100\pm24  $&$   13.77^{+0.14}_{-0.09}$ \\
  C\,IV   &   1548.20   &$-107\pm 7$&$   38\pm 6  $&$   170\pm41  $&$   13.75^{+0.14}_{-0.09}$ \\
  N\,V    &   1242.80   &$-119\pm 5$&$   52\pm10  $&$    13\pm 4    $&$   13.08^{+0.12}_{-0.15}$  \\
  N\,V    &   1238.82   &$-109\pm 7$&$   34\pm10  $&$    31\pm 8    $&$   13.16^{+0.10}_{-0.13}$  \\
  O\,VI   &   1031.93   &$-115\pm 7$&$   43\pm 7  $&$   103\pm27  $&$    14.05^{+0.16}_{-0.09}$ \\
  Si\,III   &   1206.50   &$-118\pm 6$&$   42\pm 8    $&$   355\pm67  $&$  13.74^{+0.15}_{-0.13}$ \\
  Si\,IV  &   1402.77   &$-106\pm11$&$   37\pm 5  $&$    56\pm16  $&$    13.15^{+0.15}_{-0.12}$ \\
  Si\,IV  &   1393.76   &$-109\pm 8$&$   39\pm 9   $&$    91\pm25  $&$    13.08^{+0.13}_{-0.12}$ \\
 \enddata

\tablenotetext{a}{Table lists data for measured HVC absorption lines, including ion and rest wavelength,
    HVC velocity centroid ($V_c$), line width ($b_{\rm width}$), equivalent width ($W_{\lambda}$), and column
    density ($N_a$) from apparent optical depth (AOD) integration.  Error bars are $1\,\sigma$, including both
    statistical errors in measurement and systematic effects of continuum placement and velocity range.  
    For the HVC at  $\langle V_{\rm LSR} \rangle = -108$~\kms\ our GBT measurements (Table 2) give
    $\log N_{\rm HI} = 19.50\pm0.01$.   Wakker \etal\ (2003) quote log~\NHI\ $= 19.51 \pm 0.01$.   }

\end{deluxetable}

%%%%%%%%%%%%%%%%%%%%%%%

%%%%%
%%%%%  Table 4 -- revised by Matt Stevans  (HVC data for Mrk 290 - lower-velocity HVC)  %%%%%%%%%
%%%%%    Shortened the significant figures (M. Shull, March 11)
%%%%%    combined 3a+3b, CD 3/11

\begin{deluxetable}{llllll}
\tabletypesize{\footnotesize}
\tablecaption{HVC Measurements (Mrk 290) }
\tablecolumns{6}
\tablewidth{0pt}
\tablehead{
   \colhead{Ion}   &
   \colhead{$\lambda_{0}$}   &
   \colhead{$V_c$} &
   \colhead{$b_{\rm width}$}   &
   \colhead{$W_{\lambda}$} &
   \colhead{$\log N_a$}    \\
   \colhead{}        &
   \colhead{(\AA)}   & 
   \colhead{(\kms)}  &
   \colhead{(\kms)}  &
   \colhead{(m\AA)}  &
   \colhead{($N$ in cm$^{-2}$) }    
 }

\startdata  
\cutinhead{$v_{\rm LSR} \approx-120$ \kms\ Absorber\tablenotemark{a} }
   S\,II    & 1259.51    & $-123\pm 5$  & $32\pm 5$     & $ 55\pm 6$        & $14.43^{+0.05}_{-0.04}$  \\
   S\,II    & 1253.80     & $-115\pm 8$ & $ 35\pm 10$  & $ 11\pm 2$        & $13.93^{+0.06}_{-0.07}$  \\
   S\,II    & 1250.57     & $-104\pm12$ & $34\pm 4$    & $ 83\pm35$       & $15.14^{+0.26}_{-0.19}$    \\
   Si\,II   & 1526.70     & $-123\pm 2$ & $37\pm 9$     & $317\pm64$      & $14.32^{+0.08}_{-0.06}$  \\
   Si\,II   & 1304.37     & $-118\pm 4$ & $34\pm 6$     & $207\pm33$      & $14.40^{+0.09}_{-0.04}$  \\
   Si\,II  & 1193.28     & $-114\pm 5$ & $34\pm 8$      & $269\pm59$      & $14.16^{+0.13}_{-0.07}$   \\
   Fe\,II  & 1608.45    & $-121\pm 3$ & $35\pm 7$      & $180\pm28$      & $14.26^{+0.08}_{-0.05}$   \\
   Fe\,II  & 1144.93    & $-128\pm 4$ & $38\pm 9$      & $141\pm30$      & $14.30^{+0.11}_{-0.09}$  \\
   C\,II*  & 1335.66    & $-120\pm39$ & $43\pm17$    & $<15$                &   $<14.35(3\sigma)$           \\
   C\,II   & 1334.53     & $-122\pm 3$ & $40\pm12$     & $389\pm11$     & $14.85^{+0.50}_{-0.49}$  \\
   N\,I    & 1200.71     & $-127\pm 1$ & $34\pm 7$      & $212\pm30$      & $14.82^{+0.06}_{-0.04}$  \\
   N\,I    & 1200.22    & $-135\pm 4$ & $39\pm11$      & $210\pm66$      & $14.51^{+0.19}_{-0.14}$  \\
   N\,I    & 1199.54    & $-119\pm 3$ & $36\pm 8$       & $144\pm26$      & $14.06^{+0.10}_{-0.05}$  \\
   O\,I    & 1302.16    & $-130\pm 5$ & $26\pm 4$       & $187\pm 4$        & $14.64^{+0.03}_{-0.02}$  \\
   Al\,II  & 1670.78    & $-124\pm 4$ & $38\pm10$      & $339\pm83$       & $13.11^{+0.12}_{-0.09}$  \\
   P\,II   & 1152.82    & $-120\pm25$ & $45\pm10$    & $ 16\pm 4$          & $12.83^{+0.22}_{-0.18}$  \\
   C\,IV  & 1550.78    & $-118\pm3$  & $38\pm12$    & $104\pm31$        & $13.79^{+0.13}_{-0.13}$   \\  
   C\,IV  & 1548.20    & $-115\pm5$  & $37\pm11$    & $177\pm54$        &  $13.77^{+0.15}_{-0.13}$  \\
   N\,V    & 1242.80   & $-131\pm17$ & $50\pm20$   & $  8\pm 4$            & $12.89^{+0.55}_{-0.18}$  \\
   N\,V    & 1238.82   & $-105\pm 1$ & $37\pm 7$     & $ 18\pm 7$           & $12.98^{+0.22}_{-0.15}$  \\
   O\,VI   & 1031.93  & $-124\pm 1$ & $38\pm11$    & $113\pm38$         & $14.10^{+0.15}_{-0.14}$  \\
   Si\,III & 1206.50     & $-120\pm 1$ & $40\pm12$   & $345\pm96$         & $13.73^{+0.18}_{-0.15}$  \\
   Si\,IV  & 1402.77   & $-116\pm 3$ & $39\pm11$   & $ 70\pm19$          & $13.26^{+0.16}_{-0.10}$  \\
   Si\,IV  & 1393.76   & $-120\pm 2$ & $41\pm11$   & $108\pm33$         & $13.17^{+0.15}_{-0.12}$   \\
   
 \cutinhead{$v_{\rm LSR} \approx-220$ \kms\ Absorber\tablenotemark{b}}
 
   P\,II   & 1152.82     & $-225\pm12$ & $56\pm 5$  & $<15$                &  $<13.19(3 \sigma$)      \\  
   C\,IV   & 1550.78   & $-224\pm 1$ & $33\pm11$  & $ 71\pm17$      & $13.60^{+0.10}_{-0.12}$ \\
   C\,IV   & 1548.20   & $-223\pm 4$ & $35\pm 5$   & $110\pm14$     & $13.51^{+0.08}_{-0.04}$ \\
   O\,VI   & 1031.93   & $-219\pm 8$ & $33\pm 4$  & $ 53\pm14$       & $13.75^{+0.15}_{-0.22}$ \\
   Si\,IV  & 1402.77   & $-218\pm 2$ & $33\pm12$  & $ 35\pm 8$       & $12.95^{+0.12}_{-0.13}$ \\
   Si\,IV  & 1393.76   & $-218\pm 7$ & $32\pm 7$   & $ 69\pm12$      & $12.96^{+0.07}_{-0.09}$ \\
   Si\,III & 1206.50     & $-211\pm 7$ & $31\pm 5$  & $180\pm38$     & $13.15^{+0.10}_{-0.08}$ \\
\enddata

\tablenotetext{a}{Table lists data for measured HVC absorption lines, including ion and rest wavelength,
    HVC velocity centroid ($V_c$), line width ($b_{\rm width}$), equivalent width ($W_{\lambda}$), and 
    column density ($N_a$) from apparent optical depth (AOD) integration.  Error bars are $1\,\sigma$, including 
    both statistical errors in measurement and systematic effects of continuum placement and velocity range.  
    For the composite UV absorber at $\langle V_{\rm LSR} \rangle = -120$ \kms, our GBT measurements (Table 2) 
    give $\log N_{\rm HI} = 20.05\pm0.02$ for components at $V_{\rm LSR} = -136, -113$, and
    $-90$ \kms.  Wakker \etal\ (2003) measure a combined log~\NHI\  $= 20.10\pm 0.02$ for HVCs at 
    $-134$ \kms\ and $-105$~\kms.   }

\tablenotetext{b}{For the UV-only absorber at $\langle V_{\rm LSR} \rangle -220$~\kms\, 
Wakker \etal\ (2003) find  no H\,I emission to a limit  $\log N_{\rm HI} < 18.3$.   Our GBT limit 
over the range ($-275$ to $-175$ \kms)  is $\log N_{\rm HI} < 18.0$ (at $4 \sigma$).
 }

\end{deluxetable}

%%%%%%%%%%%%%%%%%%%%%%%

\newpage

%%%%%
%%%%%  Table 5 -- revised by Matt Stevans  (HVC data for Mrk 876 - both HVCs)  %%%%%%%%%

\begin{deluxetable}{llllll}
\tabletypesize{\footnotesize}
\tablecaption{HVC Measurements\tablenotemark{a} (Mrk 876) }
\tablecolumns{6}
\tablewidth{0pt}
\tablehead{
   \colhead{Ion}   &
   \colhead{$\lambda_{0}$}   &
   \colhead{$V_c$} &
   \colhead{$b_{\rm width}$}   &
   \colhead{$W_{\lambda}$} &
   \colhead{$\log N_a$}    \\
   \colhead{}        &
   \colhead{(\AA)}   & 
   \colhead{(\kms)}  &
   \colhead{(\kms)}  &
   \colhead{(m\AA)}  &
   \colhead{($N$ in cm$^{-2}$) }    
}

\startdata  
\cutinhead{$v_{\rm LSR} \approx-133$ \kms\ Absorber\tablenotemark{a} }
   S\,II   &   1259.51   &$-132\pm3   $&$25\pm10   $&$   37\pm  13   $&$   14.24^{+ 0.16 }_{ -0.13}$ \\
   S\,II   &   1253.80   &$-133\pm2   $&$24\pm11   $&$   25\pm    8   $&$   14.25^{+ 0.13 }_{ -0.14}$ \\
   S\,II   &   1250.57   &$-137\pm3   $&$24\pm10   $&$   16\pm    5   $&$   14.37^{+ 0.14 }_{ -0.13}$ \\
   Si\,II  &   1526.70   &$-135\pm1   $&$26\pm10   $&$  259\pm  86  $&$   14.30^{+ 0.14 }_{ -0.08}$ \\
   Si\,II  &   1304.37   &$-134\pm1   $&$26\pm11   $&$  191\pm  65  $&$   14.41^{+ 0.16 }_{ -0.09}$ \\  
   Si\,II  &   1193.28   &$-134\pm1   $&$28\pm12   $&$  242\pm 100 $&$   13.97^{+ 0.19 }_{ -0.12}$ \\ 
   Si\,II  &   1190.41   &$-133\pm1   $&$27\pm12   $&$  227\pm  90   $&$   14.19^{+ 0.18 }_{ -0.11}$ \\ 
   Fe\,II  &   1608.45   &$-136\pm1   $&$26\pm11   $&$ 168\pm  56  $&$    14.26^{+ 0.17 }_{ -0.10}$ \\ 
   Fe\,II  &   1144.93   &$-134\pm2   $&$27\pm11   $&$ 118\pm  43   $&$   14.24^{+ 0.18 }_{ -0.12}$ \\
   Fe\,II  &   1143.22   &$-140\pm2  $&$24\pm 6    $&$   34\pm  6      $&$    14.25^{+ 0.13 }_{ -0.06}$ \\ 
   Fe\,II  &   1133.67   &$-132\pm4   $&$29\pm8    $&$   31\pm  13    $&$   14.83^{+ 0.27 }_{ -0.15}$ \\  
   P\,II    &   1152.82   &$-130\pm1   $&$29\pm12  $&$  24\pm  10    $&$   12.97^{+ 0.32 }_{ -0.12}$ \\
   C\,II*  &   1335.66   &$-137\pm7   $&$26\pm13   $&$   14\pm   8     $&$   13.86^{+ 0.21 }_{ -0.24}$ \\
   C\,II   &   1334.53   &$-134\pm1   $&$28\pm13   $&$  292\pm 134  $&$   14.79^{+ 0.24 }_{ -0.16}$ \\
   N\,I    &   1200.71    &$-139\pm3   $&$26\pm10   $&$  213\pm  78   $&$   14.99^{+ 0.19 }_{ -0.11}$ \\
   N\,I    &   1200.22   &$-142\pm4   $&$26\pm11    $&$  148\pm  68   $&$   14.36^{+ 0.26 }_{ -0.23}$ \\
   N\,I    &   1199.54   &$-134\pm2   $&$24\pm9      $&$  112\pm  32   $&$   13.97^{+ 0.13 }_{ -0.08}$ \\
   O\,I    &   1302.16   &$-136\pm1   $&$26\pm11    $&$  206\pm  79   $&$   14.74^{+ 0.16 }_{ -0.12}$ \\
   Al\,II   &   1670.78   &$-137\pm1   $&$27\pm12    $&$  280\pm 105  $&$  13.12^{+ 0.17 }_{ -0.11}$ \\  
   C\,IV   &   1550.78   &$-130\pm3   $&$29\pm13   $&$   73\pm  37   $&$   13.63^{+ 0.31 }_{ -0.18}$ \\
   C\,IV   &   1548.20   &$-129\pm2   $&$29\pm12   $&$  117\pm 57   $&$   13.58^{+ 0.29 }_{ -0.18}$ \\
   N\,V    &   1242.80   &$-135\pm1   $&$28\pm15   $&$    7\pm    4    $&$   12.82^{+ 0.18 }_{ -0.29}$ \\
   N\,V    &   1238.82   &$-139\pm2   $&$27\pm14   $&$   14\pm   6   $&$    12.85^{+ 0.23 }_{ -0.14}$ \\
   O\,VI   &   1031.93   &$-136\pm1   $&$30\pm13   $&$   86\pm  43   $&$   13.99^{+ 0.36 }_{ -0.14}$ \\ 
   Si\,IV  &   1402.77   &$-128\pm4   $&$28\pm11   $&$   48\pm  20   $&$   13.09^{+ 0.24 }_{ -0.14}$ \\
   Si\,IV  &   1393.76   &$-129\pm3   $&$24\pm11   $&$   91\pm  39   $&$   13.10^{+ 0.25 }_{ -0.15}$ \\ 
   Si\,III &   1206.50   &$-134\pm1   $&$29\pm12   $&$  269\pm 117   $&$   13.72^{+ 0.24 }_{ -0.14}$ \\
     
\cutinhead{$v_{\rm LSR} \approx-190$ \kms\ Absorber\tablenotemark{b}}

  Si\,III   &   1206.50    &$   -198\pm   8   $&$   27\pm  7   $&$  162\pm 59    $&$   13.17^{+0.21}_{-0.19}$ \\
  Si\,II    &   1526.70    &$   -197\pm   8   $&$   24\pm  6   $&$  129\pm 42    $&$   13.81^{+0.13}_{-0.18}$ \\
  Si\,II    &   1304.37    &$   -200\pm   8   $&$   31\pm  6   $&$    87\pm 30    $&$   13.92^{+0.17}_{-0.15}$ \\
  Si\,II    &   1193.28    &$   -196\pm   8   $&$   25\pm  5   $&$  145\pm 49    $&$   13.56^{+0.21}_{-0.13}$ \\
  Si\,II    &   1190.41    &$   -196\pm   7   $&$   27\pm  7   $&$  118\pm 42    $&$   13.69^{+0.21}_{-0.14}$ \\
  Fe\,II   &   1608.45    &$   -200\pm   9   $&$   30\pm  9   $&$   77\pm 32     $&$    13.84^{+0.17}_{-0.20}$ \\
  Fe\,II   &   1144.93    &$   -200\pm   4   $&$   33\pm 14  $&$   40\pm 21     $&$   13.68^{+0.22}_{-0.23}$ \\
   O\,I     &   1302.16    &$   -199\pm   7   $&$   28\pm  6   $&$  165\pm 47    $&$   14.57^{+0.15}_{-0.12}$ \\
  Al\,II    &   1670.78    &$   -202\pm 9    $&$   30\pm  6    $&$  151\pm 54    $&$   12.64^{+0.14}_{-0.22}$ \\
  C\,IV   &   1550.78    &$   -191\pm 14  $&$   18\pm  3    $&$   25\pm 12     $&$   13.13^{+0.15}_{-0.12}$ \\
  C\,IV   &   1548.20    &$   -192\pm  10  $&$   27\pm  5   $&$   43\pm 19    $&$    13.09^{+0.25}_{-0.17}$ \\

  Si\,IV  &   1402.77    &$   -188\pm  20  $&$   13\pm  8    $&$   11\pm  3     $&$    12.40^{+0.25}_{-0.12}$ \\
  Si\,IV  &   1393.76    &$   -195\pm   6   $&$   34\pm 13   $&$   22\pm 13    $&$   12.44^{+0.31}_{-0.24}$ \\

\enddata

\tablenotetext{a}{Table lists data for measured HVC absorption lines, including ion and rest wavelength,
    HVC velocity centroid ($V_c$), line width ($b_{\rm width}$), equivalent width ($W_{\lambda}$), and column
     density ($N_a$) from apparent optical depth (AOD) integration.   Error bars are $1\,\sigma$, including both
    statistical errors in measurement and systematic effects of continuum placement and velocity range.  
    Our GBT measurements give $\log N_{\rm HI} = 19.39\pm0.01$ and  $\log N_{\rm HI} = 18.71\pm0.04$ for 
    HVC components at $V_{\rm LSR} = -131$ \kms\ and $-175$ \kms, respectively (Table 2).  Wakker \etal\ (2003) 
    measure log~\NHI\  $= 19.30\pm 0.02$ at $V_{\rm LSR} = -133$~\kms.   }
      
\tablenotetext{b}{For this UV-only HVC, our GBT measurements (Table 2) give
    $\log N_{\rm HI} = 18.71\pm0.04$ at $V_{\rm LSR} = -175$ \kms.  
    Wakker \etal\ (2003) report no H\,I emission at $-190$~\kms, but  they find
    log~\NHI\  $= 18.67\pm 0.08$ for a weak component at $V_{\rm LSR} = -173$~\kms. }

\end{deluxetable}

%%%%%%%%%%%%%%%%%%%%%%%

%%%%%
%%%%%  Table 6 -- revised by Matt Stevans  (HVC data for PG 1259)  %%%%%%%%%

\begin{deluxetable}{llllll}
\tabletypesize{\footnotesize}
\tablecaption{HVC Measurements\tablenotemark{a} (PG\,1259$+$593) }
\tablecolumns{6}
\tablewidth{0pt}
\tablehead{
   \colhead{Ion}   &
   \colhead{$\lambda_{0}$}   &
   \colhead{$V_c$} &
   \colhead{$b_{\rm width}$}   &
   \colhead{$W_{\lambda}$} &
   \colhead{$\log N_a$}    \\
   \colhead{}        &
   \colhead{(\AA)}   & 
   \colhead{(\kms)}  &
   \colhead{(\kms)}  &
   \colhead{(m\AA)}  &
   \colhead{($N$ in cm$^{-2}$) }    
 }

%\tableline
\startdata  
  S\,II    &   1259.51   &$   -127\pm   6   $&$   23\pm   7    $&$   38\pm  10     $&$   14.27^{+0.11}_{-0.11}$ \\
  S\,II    &   1253.80   &$   -128\pm   5   $&$   25\pm   9    $&$   30\pm   8      $&$   14.35^{+0.15}_{-0.10}$ \\
  S\,II    &   1250.57   &$   -120\pm   8   $&$   18\pm   1    $&$   22\pm   5      $&$   14.51^{+0.11}_{-0.10}$ \\
  Si\,II   &   1526.70   &$   -124\pm   6   $&$   25\pm   9     $&$  188\pm  62   $&$   14.06^{+0.14}_{-0.15}$ \\
  Si\,II   &   1304.37   &$   -124\pm   6   $&$   22\pm   8     $&$  126\pm  39   $&$   14.17^{+0.10}_{-0.14}$ \\
  Si\,II   &   1193.28   &$   -121\pm   7   $&$   24\pm   7     $&$  166\pm  55   $&$   13.66^{+0.19}_{-0.15}$ \\
  Si\,II   &   1190.41   &$   -122\pm   5   $&$   25\pm   9     $&$  160\pm  56   $&$   13.90^{+0.17}_{-0.14}$ \\
  Fe\,II   &   1608.45   &$   -126\pm   7   $&$   24\pm   7     $&$  123\pm  34   $&$   14.11^{+0.12}_{-0.11}$ \\
  Fe\,II   &   1144.93   &$   -126\pm   6   $&$   29\pm  12    $&$   97\pm  39    $&$   14.13^{+0.22}_{-0.16}$ \\
  Fe\,II   &   1143.22   &$   -138\pm   0   $&$   19\pm  14   $&$   26\pm   5       $&$   14.16^{+0.11}_{-0.12}$ \\
  P\,II     &   1152.82   &$   -130\pm   3    $&$  13 \pm28     $&$  <10                 $&$   <13.02(3\sigma)$   \\
  C\,II*   &   1335.66   &$   -140\pm  13   $&$   33\pm   8    $&$   35\pm  24     $&$   14.29^{+0.34}_{-0.31}$ \\
  C\,II    &   1334.53   &$   -122\pm   5    $&$    26\pm   9     $&$  219\pm  75   $&$   14.45^{+0.23}_{-0.17}$ \\
  N\,I      &   1200.71   &$   -130\pm   1    $&$   30\pm  12    $&$  193\pm  77   $&$   14.79^{+0.21}_{-0.14}$ \\
  N\,I      &   1200.22   &$   -140\pm   3    $&$   27\pm  12    $&$  167\pm  67   $&$   14.44^{+0.23}_{-0.14}$ \\
  N\,I      &   1199.54   &$   -127\pm   5    $&$   24\pm   9     $&$   89\pm  26    $&$   13.85^{+0.12}_{-0.12}$ \\
  Al\,II    &   1670.78   &$   -127\pm   5    $&$   26\pm   9     $&$  214\pm  69   $&$   12.89^{+0.13}_{-0.16}$ \\
  C\,IV   &   1550.78   &$   -124\pm   9   $&$   27\pm   8      $&$   25\pm  10     $&$   13.13^{+0.24}_{-0.16}$ \\
  C\,IV   &   1548.20   &$   -115\pm   8   $&$   23\pm   7      $&$   41\pm  19     $&$   13.08^{+0.33}_{-0.15}$ \\
  N\,V    &   1242.80   &$   -125\pm   6   $&$   31\pm  13    $&$   <14                 $&$   <13.59(3\sigma)$  \\
  N\,V    &   1238.82   &$   -122\pm   9   $&$   23\pm  10    $&$    <7                  $&$    <12.99(3\sigma)$   \\
  O\,VI   &   1031.93   &$   -130\pm   3   $&$   27\pm   8     $&$   39\pm  13      $&$   13.58^{+0.10}_{-0.12}$ \\
  Si\,IV  &   1402.77   &$   -119\pm   4    $&$   25\pm  11   $&$   23\pm   9       $&$   12.75^{+0.22}_{-0.16}$ \\
  Si\,IV  &   1393.76   &$   -124\pm   6    $&$   26\pm   8    $&$   34\pm  12      $&$   12.64^{+0.17}_{-0.13}$ \\
  Si\,III &     1206.50   &$   -124\pm   6     $&$   27\pm   8     $&$  187\pm  67     $&$   13.30^{+0.24}_{-0.16}$ \\
 \enddata

\tablenotetext{a}{Table lists data for measured HVC absorption lines, including ion and rest wavelength,
    HVC velocity centroid ($V_c$), line width ($b_{\rm width}$), equivalent width ($W_{\lambda}$), and column
    density ($N_a$) from apparent optical depth (AOD) integration.   Error bars are $1\,\sigma$, including both
    statistical errors in measurement and systematic effects of continuum placement and velocity range.  
    The GBT measurements give $\log N_{\rm HI} = 19.97 \pm 0.02$ for the two-component HVC ($-126.4$~\kms\
    and $=129.1$~\kms) at  $\langle V_{\rm LSR} \rangle = -128$ \kms.   Wakker \etal\ (2003) measure 
     log~\NHI\  $= 19.95\pm 0.01$ for an HVC at $V_{\rm LSR} = -128$~\kms, and  Sembach \etal\ (2004)
    adopt log~\NHI\  $= 19.94\pm 0.06$ for this HVC. }

\end{deluxetable}

%%%%%%%%%%%%%%%%%%%%%%%

%%%%%  Table 7 (Summary of Columns and Abundances) - Mrk 817   %%%%%%%%%

\begin{deluxetable}{llllll}
\tabletypesize{\footnotesize}
\tablecaption{Mrk~817:  Summary Column Densities\tablenotemark{a} and Abundances}
\tablecolumns{6}
\tablewidth{0pt}
\tablehead{
\colhead{Species}  &  \colhead{CSG-03}           &  \colhead{Shull-11}    &  \colhead{Adopted}   
     &   \colhead{Abundance\tablenotemark{b}}  &    \colhead{Abundance\tablenotemark{b}}  \\
       $(X_i)$      &   ($\log N_{X_i}$) &  ($\log N_{X_i}$)  &  
       ($\log N_{X_i}$)  &     $\log (N_{X_i}/N_{\rm HI})$ & $[X/H]$   
       }   
\startdata    
O\,I                    &   $15.72^{+0.24}_{-0.16}$         &  $\geq14.98$                              
                          &    $15.72^{+0.24}_{-0.16}$        &  $-3.78\pm0.24$       &  $-0.47 \pm 0.24$    \\          
N\,I                    &    $<14.05$                                   & $14.06^{+0.12}_{-0.07}$          
                          &    $14.06^{+0.12}_{-0.07}$       &  $-5.44\pm0.10$        & $-1.27 \pm 0.10$     \\          
S\,II                  &   $14.34^{+0.05}_{-0.05}$          &  $14.29^{+0.08}_{-0.08}$        
                         &   $14.29^{+0.08}_{-0.08}$          &   $-5.21\pm0.08$    &  $-0.60 \pm 0.10$  \\                   
Si\,II                 &  $14.48^{+0.07}_{-0.08}$           &  $14.47^{+0.08}_{-0.05}$          
                         &   $14.47^{+0.08}_{-0.05}$          &  $-5.03\pm0.08$     &   $-0.75 \pm 0.10$   \\ 
Fe\,II                &  $14.31^{+0.11}_{-0.08}$            &  $14.23^{+0.10}_{-0.06}$          
                         &      $14.23^{+0.10}_{-0.06}$       &  $-5.27\pm0.10$     &  $-0.88 \pm 0.15$   \\      
C\,II                  &  $\cdots$                                        &  $\geq14.9$                                 
                         &     $\geq14.9$                               &   $>-4.6$                    &  $\geq -1.03  $  \\                                   
Al\,II                  &     $\cdots$                                    & $13.23^{+0.07}_{-0.06}$           
                          & $13.23^{+0.07}_{-0.06}$          &   $-6.27\pm0.07$      &  $- 0.72 \pm 0.07$   \\             
P\,II                   &     $\cdots$                                    &  $<12.95$                                    
                          &    $<12.95$                                   &   $<-6.55$                  &  $<0.04$   \\          
C\,IV                 &     $\cdots$                                    & $13.77^{+0.14}_{-0.09}$           
                          &    $13.77^{+0.14}_{-0.09}$       &  $-5.73^{+0.14}_{-0.09}$    &     \\          
N\,V                  &     $\cdots$                                    &   $13.16^{+0.10}_{-0.13}$                               
                          &    $13.16^{+0.10}_{-0.13}$        & $-6.34^{+0.10}_{-0.13}$      &     \\          
O\,VI                 &     $\cdots$                                    &  $14.05^{+0.16}_{-0.09}$        
                         &  $14.05^{+0.16}_{-0.09}$          & $-5.45^{+0.16}_{-0.09}$      &       \\          
Si\,IV                &    $\cdots$                                     &  $13.08^{+0.13}_{-0.12}$        
                         &  $13.08^{+0.13}_{-0.12}$          &  $-6.42^{+0.13}_{-0.12}$     &    \\      
Si\,III                 &    $\cdots$                                     &   $13.74^{+0.15}_{-0.13}$         
                          &  $13.74^{+0.15}_{-0.13}$         &  $-5.76^{+0.15}_{-0.13}$     &      \\ 
\enddata

 \tablenotetext{a}{Comparison of measured column densities, $\log N_a({\rm cm}^{-2}$), for HVC at 
 $\langle V_{\rm LSR} \rangle = -109$~\kms\ (integrated from $-190$~\kms\ to $-70$ \kms). }  

 \tablenotetext{b}{Abundances for neutrals and first ions $X_i$ are given relative to measured
 $\log N_{\rm HI} = 19.50 \pm 0.01$ (Table 2).  Fox \etal\ (2004) found 
 log~N(O\,VI) = $13.97^{+0.10}_{-0.13}$ for HVC absorption between $-160$ and $-80$~\kms.  
 Last column gives abundances of elements $[X/H]$ relative to solar abundances (Asplund \etal\ 2009),
 with ionization corrections of $0.27$ dex (S\,II), $0.21$ dex (Si\ II), and 
  $0.11$ dex (Fe\,II) subtracted from the first-ion abundances (Section 3.2). }

 \end{deluxetable}

%%%%%%%%%%%%%%%%%%%%%%%

%%%%%  Table 8 (Summary of Columns and Abundances) - Mrk 290   %%%%%%%%%

\begin{deluxetable}{lllllll}
\tabletypesize{\footnotesize}
\tablecaption{Mrk~290:  Summary Column Densities\tablenotemark{a} and Abundances}
\tablecolumns{7}
\tablewidth{0pt}
\tablehead{
\colhead{Species}  &  \colhead{CSG-03}        &  \colhead{CSG-07}  &  \colhead{Shull-11}    
     &  \colhead{Adopted}   &  \colhead{Abundance\tablenotemark{b}} &   \colhead{Abundance\tablenotemark{b}}  \\
       $(X_i)$      &  ($\log N_{X_i}$)  &  ($\log N_{X_i}$)  &   ($\log N_{X_i}$)  &  
       ($\log N_{X_i}$)     &     $\log (N_{X_i}/N_{\rm HI})$   & $[X/H]$   
       }   
\startdata    
O\,I      &   $<16.79$                                   &   $15.75^{+0.29}_{-0.15}$        &   $14.64^{+0.03}_{-0.02}$         
            &    $15.75^{+0.29}_{-0.15}$       &   $-4.30^{+0.29}_{-0.15}$         &   $-0.99^{+0.29}_{-0.15}$    \\          
N\,I      &    $<15.02$                                  &   $14.23^{+0.17}_{-0.20}$        &  $14.06^{+0.12}_{-0.07}$          
            &    $14.06^{+0.12}_{-0.07}$       &   $-5.99^{+0.12}_{-0.07}$         &   $-1.82^{+0.12}_{-0.07}$     \\          
S\,II     &  $14.29^{+0.15}_{-0.14}$         &     $14.24^{+0.16}_{-0.20}$      &   $14.43^{+0.08}_{-0.08}$         
            &  $14.43^{+0.08}_{-0.08}$         &     $-5.62^{+0.08}_{-0.08}$       &   $-0.78^{+0.15}_{-0.15}$    \\                   
Si\,II    &  $<14.97$                                    &     $14.93^{+0.18}_{-0.13}$      &   $>14.40$       
            &  $14.93^{+0.18}_{-0.13}$         &     $-5.12^{+0.18}_{-0.13}$       &   $-0.67^{+0.15}_{-0.15}$   \\ 
Fe\,II   &  $<15.46$                                    &     $14.41^{+0.23}_{-0.21}$      &   $14.26^{+0.09}_{-0.05}$         
            &  $14.26^{+0.09}_{-0.05}$         &    $-5.79^{+0.09}_{-0.05}$        &   $-1.31^{+0.09}_{-0.05}$    \\      
Al\,II    &     $\cdots$                                   &    $\cdots$                                    &   $13.11^{+0.12}_{-0.09}$         
            &  $13.11^{+0.12}_{-0.09}$         &   $-6.94^{+0.12}_{-0.09}$         &   $- 1.39^{+0.12}_{-0.09}$   \\             
P\,II     &     $\cdots$                                   &    $<12.82$                                  &   $12.83^{+0.22}_{-0.18}$         
            &    $12.83^{+0.22}_{-0.18}$       &   $-7.22^{+0.22}_{-0.18}$         &   $-0.63^{+0.22}_{-0.18}$  \\          
C\,IV   &     $\cdots$                                   &      $\cdots$                                  &  $13.78^{+0.14}_{-0.09}$          
            &    $13.78^{+0.14}_{-0.09}$       &  $-6.27^{+0.14}_{-0.09}$          &    \\          
N\,V    &     $\cdots$                                   &   $<13.49$                                    &  $12.98^{+0.22}_{-0.15}$         
            &   $12.98^{+0.22}_{-0.15}$        &   $-7.07^{+0.22}_{-0.15}$          &   \\          
O\,VI   &     $\cdots$                                   &   $14.23^{+0.04}_{-0.04}$         &    $14.10^{+0.15}_{-0.14}$       
            &   $14.10^{+0.15}_{-0.14}$         &   $-5.95^{+0.15}_{-0.14}$         &  \\          
Si\,IV   &    $\cdots$                                    &   $\cdots$                                      &    $12.95^{+0.12}_{-0.12}$       
            &   $12.95^{+0.12}_{-0.12}$         &  $-7.10^{+0.12}_{-0.12}$           &    \\       
Si\,III    &    $\cdots$                                    &  $\cdots$                                       &    $13.15^{+0.10}_{-0.08}$       
            &   $13.15^{+0.10}_{-0.08}$         &  $-6.90^{+0.10}_{-0.08}$           &   \\ 
 \enddata
 
 \tablenotetext{a}{Comparison of measured column densities, $\log N_a({\rm cm}^{-2}$), for HVC
 at $\langle V_{\rm LSR} \rangle = -120$~\kms\ (integrated from $-175$~\kms\ to $-75$~\kms). }  

 \tablenotetext{b}{Abundances for neutrals and first ions $X_i$ are given relative to measured
  $\log N_{\rm HI} = 20.05 \pm 0.02$ for all three HVC components (Table 2).   
 Wakker \etal\ (1999) found log~N(S\,II) =  $14.34^{+0.08}_{-0.11}$.  
 Last column gives elemental abundances $[X/H]$ relative to solar values (Asplund \etal\ 2009) 
 with ionization corrections of $0.04$ dex (S\,II), $0.04$ dex (Si\ II), and $0.02$ dex (Fe\,II) subtracted
 from the first-ion abundances (Section 3.2). } 
 
 \end{deluxetable}

%%%%%%%%%%%%%%%%%%%%%%%

%%%%%  Table 9 (Summary of Columns and Abundances) - Mrk 876   %%%%%%%%%

\begin{deluxetable}{lllllll}
\tabletypesize{\footnotesize}
\tablecaption{Mrk~876:  Summary Column Densities\tablenotemark{a} and Abundances}
\tablecolumns{7}
\tablewidth{0pt}
\tablehead{
\colhead{Species}  &  \colhead{CSG-03}        &  \colhead{CSG-07}    &  \colhead{Shull-11}    &  \colhead{Adopted}   
       &   \colhead{Abundance\tablenotemark{b}}    &       \colhead{Abundance\tablenotemark{b}}   \\
       $(X_i)$      &  ($\log N_{X_i}$)   &  ($\log N_{X_i}$) &  ($\log N_{X_i}$)  &  
        ($\log N_{X_i}$)  &     $\log (N_{X_i}/N_{\rm HI})$   &     $[X/H]$        }   
\startdata    
O\,I                   &   $15.55^{+0.42}_{-0.28}$        &   $15.26^{+0.17}_{-0.14}$      &   $>14.74$                                      
                         &    $15.26^{+0.17}_{-0.14}$       &   $-4.13^{+0.17}_{-0.14}$       &   $-0.82^{+0.17}_{-0.14}$   \\          
N\,I                   &   $14.20^{+0.15}_{-0.14}$        &   $13.90^{+0.03}_{-0.04}$      &   $13.97^{+0.13}_{-0.08}$          
                         &    $13.97^{+0.13}_{-0.08}$       &   $-5.42^{+0.13}_{-0.08}$        &   $-1.25^{+0.13}_{-0.08}$    \\         
S\,II                  &  $\cdots$                                      &     $<14.34$                                &   $14.25^{+0.13}_{-0.14}$         
                         &    $14.25^{+0.13}_{-0.14}$       &    $-5.14^{+0.13}_{-0.14}$       &   $-0.58^{+0.20}_{-0.20}$     \\                   
Si\,II                 &   $14.53^{+0.11}_{-0.13}$        &     $14.56^{+0.11}_{-0.09}$     &   $14.41^{+0.16}_{-0.09}$         
                         &    $14.41^{+0.16}_{-0.09}$       &   $-4.98^{+0.13}_{-0.14}$        &   $-0.76^{+0.20}_{-0.20}$     \\ 
Fe\,II                &  $14.41^{+0.09}_{-0.08}$         &    $14.36^{+0.07}_{-0.07}$      &   $14.26^{+0.17}_{-0.10}$         
                         &    $14.26^{+0.17}_{-0.10}$      &    $-5.13^{+0.17}_{-0.10}$       &   $-0.77^{+0.20}_{-0.20}$    \\      
C\,II                  &  $\cdots$                                      &     $\cdots$                                  &   $>14.76$                                    
                         &    $>14.79$                                  &   $>-4.60$                                   &   $> -1.03$      \\                                   
Al\,II                 &     $\cdots$                                   &   $13.43^{+0.31}_{-0.19}$       &   $13.12^{+0.07}_{-0.11}$          
                         &    $13.12^{+0.17}_{-0.11}$       &   $-6.27^{+0.17}_{-0.11}$        &   $-0.72^{+0.20}_{-0.20}$     \\             
P\,II                  &     $<13.21$                                 &    $\cdots$                                   &   $<12.97$         
                         &    $<12.97$                                  &   $<-6.42$                                   &   $<0.13$ \\           
C\,IV                &     $\cdots$                                   &    $13.80^{+0.04}_{-0.03}$      &   $13.58^{+0.29}_{-0.18}$           
                         &    $13.58^{+0.29}_{-0.18}$       &  $-5.81^{+0.29}_{-0.18}$         &  \\          
N\,V                 &     $\cdots$                                   &     $<13.32$                                &    $12.85^{+0.23}_{-0.14}$        
                         &    $12.85^{+0.23}_{-0.14}$       &  $-6.54^{+0.23}_{-0.14}$         &     \\
O\,VI                &    $\cdots$                                    &    $14.20^{+0.02}_{-0.02}$       &   $13.99^{+0.36}_{-0.14}$        
                         &    $13.99^{+0.36}_{-0.14}$       & $-5.40^{+0.36}_{-0.14}$           &    \\          
Si\,IV                &    $\cdots$                                    &    $13.28^{+0.03}_{-0.02}$      &   $13.10^{+0.25}_{-0.15}$       
                         &     $13.10^{+0.25}_{-0.15}$      &  $-6.29^{+0.25}_{-0.15}$          &    \\      
Si\,III                &    $\cdots$                                    &    $\cdots$                                    &    $13.72^{+0.24}_{-0.14}$        
                         &   $13.72^{+0.24}_{-0.14}$        &  $-5.67^{+0.24}_{-0.14}$          &         \\ 
\enddata

 \tablenotetext{a}{Comparison of measured column densities, $\log N_a({\rm cm}^{-2}$), for HVC 
 at $\langle V_{\rm LSR} \rangle = -133$~\kms\ (integrated from $-170$~\kms\ to $-100$~\kms).   }  
  
  \tablenotetext{b}{Abundances for neutrals and first ions $X_i$ are given relative to measured
  $\log N_{\rm HI} = 19.39 \pm 0.01$ for the $-131$~\kms\ component (Table 2). 
  Fox \etal\ (2004) found log~N(O\,VI) =  $14.12^{+0.11}_{-0.13}$ and log~N(N\,V) $< 13.43$.  
  Our previous Si\,III survey (Shull \etal\ 2009) measured log~N(Si\,III) $\geq 13.92$ as adopted here.
  Last column gives elemental abundances $[X/H]$ relative to solar values (Asplund \etal\ 2009),
  with ionization corrections of $0.32$ dex (S\,II), $0.27$ dex (Si\ II), and 
  $0.14$ dex (Fe\,II) subtracted from the first-ion abundances (Section 3.2). }

 \end{deluxetable}

%%%%%%%%%%%%%%%%%%%%%%%

%%%%%  Table 10 (Summary of Columns and Abundances) -  PG1259   %%%%%%%%%

\begin{deluxetable}{lllllll}
\tabletypesize{\footnotesize}
\tablecaption{PG1259+593:  Summary Column Densities and Abundances\tablenotemark{a}}
\tablecolumns{7}
\tablewidth{0pt}
\tablehead{
\colhead{Species}  &  \colhead{CSG-03}   & \colhead{Richter-01}  &  \colhead{Shull-11}    &  \colhead{Adopted}   
           & \colhead{Abundance\tablenotemark{b}}    &       \colhead{Abundance\tablenotemark{b}}    \\
       $(X_i)$      &  ($\log N_{X_i}$) &  ($\log N_{X_i}$)  &  ($\log N_{X_i}$)  &   ($\log N_{X_i}$)  &
      $\log (N_{X_i}/N_{\rm HI})$   &     $[X/H]$    
       }   
\startdata    
O\,I                   &   $15.75^{+0.18}_{-0.24}$         &   $15.77^{+0.37}_{-0.31}$        &  $\cdots$                                       
                         &   $15.85^{+0.15}_{-0.15}$         &   $-4.12^{+0.15}_{-0.15}$         &  $-0.81^{+0.15}_{-0.15}$      \\          
N\,I                   &   $14.02^{+0.19}_{-0.12}$         &    $13.95^{+0.17}_{-0.21}$       & $13.85^{+0.12}_{-0.12}$           
                         &   $13.85^{+0.12}_{-0.12}$         &    $-6.12^{+0.12}_{-0.12}$        & $-1.95^{+0.12}_{-0.12}$    \\          
S\,II                  &   $14.38^{+0.12}_{-0.11}$         &  $14.34^{+0.12}_{-0.15}$         &  $14.35^{+0.15}_{-0.10}$          
                         &   $14.35^{+0.15}_{-0.10}$         &   $-5.62^{+0.15}_{-0.10}$         &  $-0.78^{+0.15}_{-0.10}$\\                     
Si\,II                 &   $14.67^{+0.20}_{-0.15}$         &   $14.56^{+0.28}_{-0.27}$        &  $14.17^{+0.10}_{-0.14}$          
                         &   $14.17^{+0.10}_{-0.14}$         &   $-5.80^{+0.10}_{-0.14}$           &  $-1.35^{+0.10}_{-0.14}$\\ 
Fe\,II                 &  $14.40^{+0.17}_{-0.08}$         &    $14.16^{+0.20}_{-0.14}$       &  $14.11^{+0.12}_{-0.11}$          
                         &   $14.11^{+0.12}_{-0.11}$         &   $-5.86^{+0.12}_{-0.11}$         &  $-1.38^{+0.12}_{-0.11}$   \\      
C\,II                  &  $\cdots$                                       &  $\cdots$                                      &  $>14.45$                  
                         &    $>14.45$                                   & $>-5.52$                                      &  $>-1.95$      \\                                   
Al\,II                  &  $13.45^{+0.17}_{-0.16}$          &   $13.42^{+0.30}_{-0.50}$       & $12.89^{+0.13}_{-0.16}$      
                         &  $12.89^{+0.13}_{-0.16}$          &  $-7.08^{+0.13}_{-0.16}$         &    $-1.53^{+0.13}_{-0.16}$  \\             
P\,II                  &     $<12.96$                                   &    $<13.22$                                 &  $<12.43$                                     
                         &   $<12.43$                                    &   $<-7.54$                                    &  $<-0.95$          \\          
C\,IV                &     $\cdots$                                    &   $\cdots$                                     & $13.08^{+0.33}_{-0.15}$           
                         &   $13.08^{+0.33}_{-0.15}$        &   $-6.89^{+0.33}_{-0.15}$          &     \\          
N\,V                 &     $\cdots$                                    &    $\cdots$                                    &  $<12.99$                                      
                         &   $<12.99$                                    &  $<-6.98$                                     &      \\          
O\,VI                 &     $\cdots$                                   &  $\cdots$                                      &    $13.58^{+0.18}_{-0.12}$        
                         &  $13.58^{+0.18}_{-0.12}$          &   $-6.39^{+0.18}_{-0.12}$         &   \\          
Si\,IV                &    $\cdots$                                     &  $\cdots$                                      &  $12.64^{+0.17}_{-0.13}$   
                         &   $12.64^{+0.17}_{-0.13}$          &  $-7.33^{+0.17}_{-0.13}$         &  \\      
Si\,III                &    $\cdots$                                      & $\cdots$                                       &   $13.30^{+0.24}_{-0.16}$         
                         & $13.30^{+0.24}_{-0.16}$           &  $-6.67^{+0.24}_{-0.16}$           &        \\ 
\enddata

 \tablenotetext{a}{Comparison of measured column densities, $\log N_a({\rm cm}^{-2}$), for HVC at
 $\langle V_{\rm LSR} \rangle = -128$~\kms\ (integrated from $-170$ to $-95$ \kms).   }

  \tablenotetext{b}{Abundances for neutrals and first ions $X_i$ are given relative to measured
  $\log N_{\rm HI} = 19.97 \pm 0.02$ for the HVC components at $-126$~\kms\ and $-129$ \kms\ 
  (Table 2).   We adopted $\log N_{\rm OI} = 15.85 \pm 0.15$ from Sembach \etal\ (2004).  
  Fox \etal\ (2004) found log~N(C\,IV) $= 13.26^{+0.04}_{-0.06}$, log~N(N\,V) $<12.85$,
 log~N(Si\,IV) $= 12.73^{+0.05}_{-0.03}$,  log~N(O\,VI) $= 13.71^{+0.09}_{-0.09}$.  Our previous
 Si\,III survey (Shull \etal\ 2009) measured log~N(Si\,III) $\geq13.60$ as adopted here.  
 Last column gives elemental abundances $[X/H]$ relative to solar values (Asplund \etal\ 2009),
 with ionization corrections of $0.04$ dex (S\,II), $0.04$ dex (Si\ II), and 
  $0.02$ dex (Fe\,II) subtracted from the first-ion abundances (Section 3.2). }

 \end{deluxetable}

%%%%%%%%%%%%%%%%%%%%%%%

%%%%%%%%%%%%%%%%%%%%%%%%%%%%%%%%%%%%%%%%%%%%%%%%%%
%% Figure 4

\begin{figure}
   \epsscale{0.9} 
   \plotone{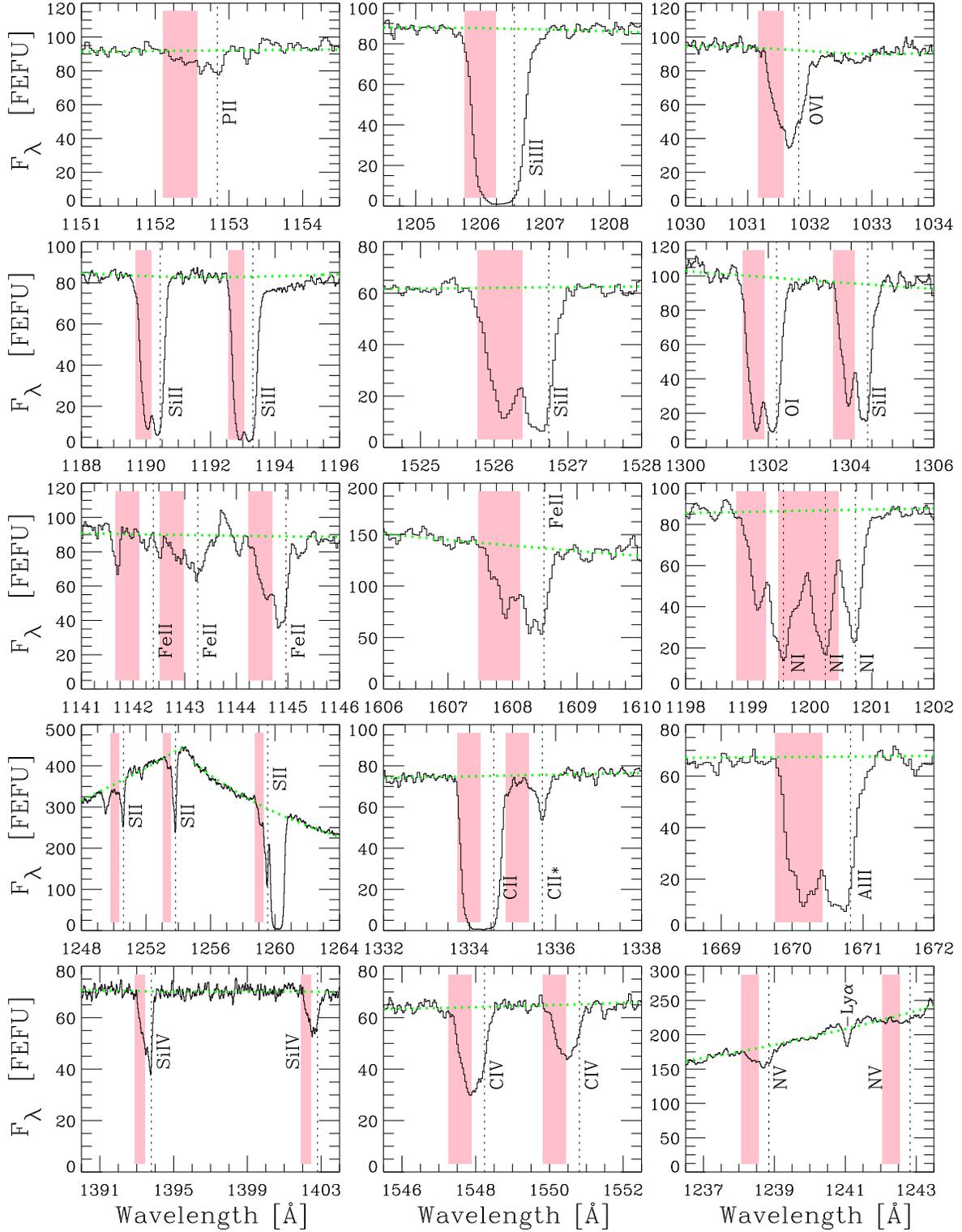}    
  \caption{ Detail of COS (G130M, G160M) data on HVC at $\langle V_{\rm LSR} \rangle = -109$ \kms\ 
   (range $-190$ to $-70$ \kms) toward Mrk~817, showing absorption lines of both low and high ions.   
   HVC Complex~C is seen in UV absorption indicated by pink wash.  
   Fluxes are shown in FEFU (femto-erg flux units) or $10^{-15}$ erg cm$^{-2}$ s$^{-1}$ \AA$^{-1}$. 
      }
\end{figure} 
  
%%%%%%%%%%%%%%%%%%%%%%%%%%%%%%%%%%%%%%%%%%%%%%%%

%%%%%%%%%%%%%%%%%%%%%%%%%%%%%%%%%%%%%%%%%%%%%%%%%%
%% Figure 5

\begin{figure}
   \epsscale{0.9} 
   \plotone{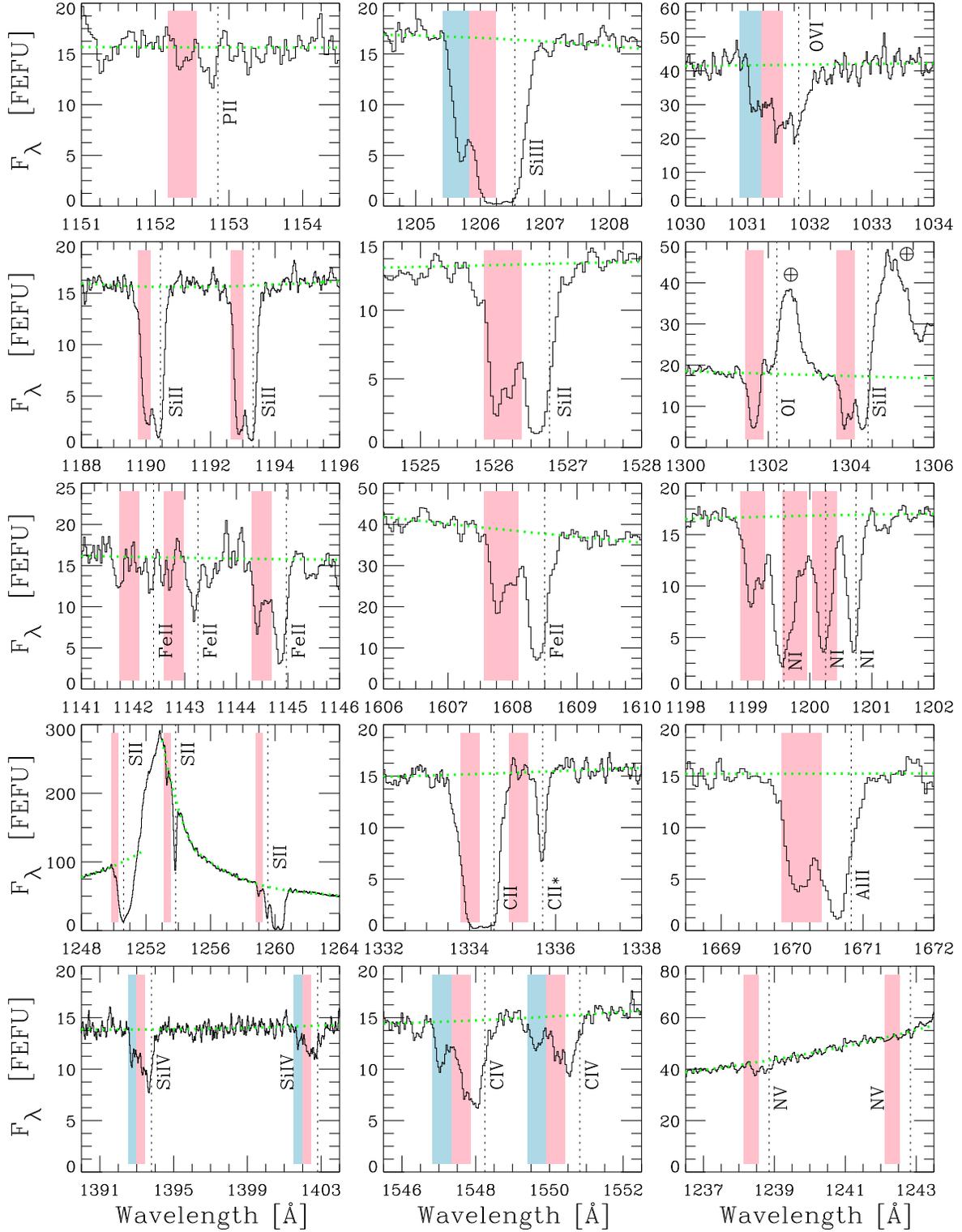}    
   \caption{Same as Fig.\ 4 for Mrk~290 sight line.  Note two HVCs shown by pink and blue wash,
     corresponding to components at $\langle V_{\rm LSR} \rangle = -120$ \kms\ (range $-175$ to 
     $-75$ \kms) and $\langle V_{\rm LSR} \rangle =-220$ \kms\ (range $-275$ to $-175$ \kms).
      }
\end{figure} 
  
%%%%%%%%%%%%%%%%%%%%%%%%%%%%%%%%%%%%%%%%%%%%%%%%

%%%%%%%%%%%%%%%%%%%%%%%%%%%%%%%%%%%%%%%%%%%%%%%%%%
%% Figure 6

\begin{figure}
   \epsscale{0.9} 
   \plotone{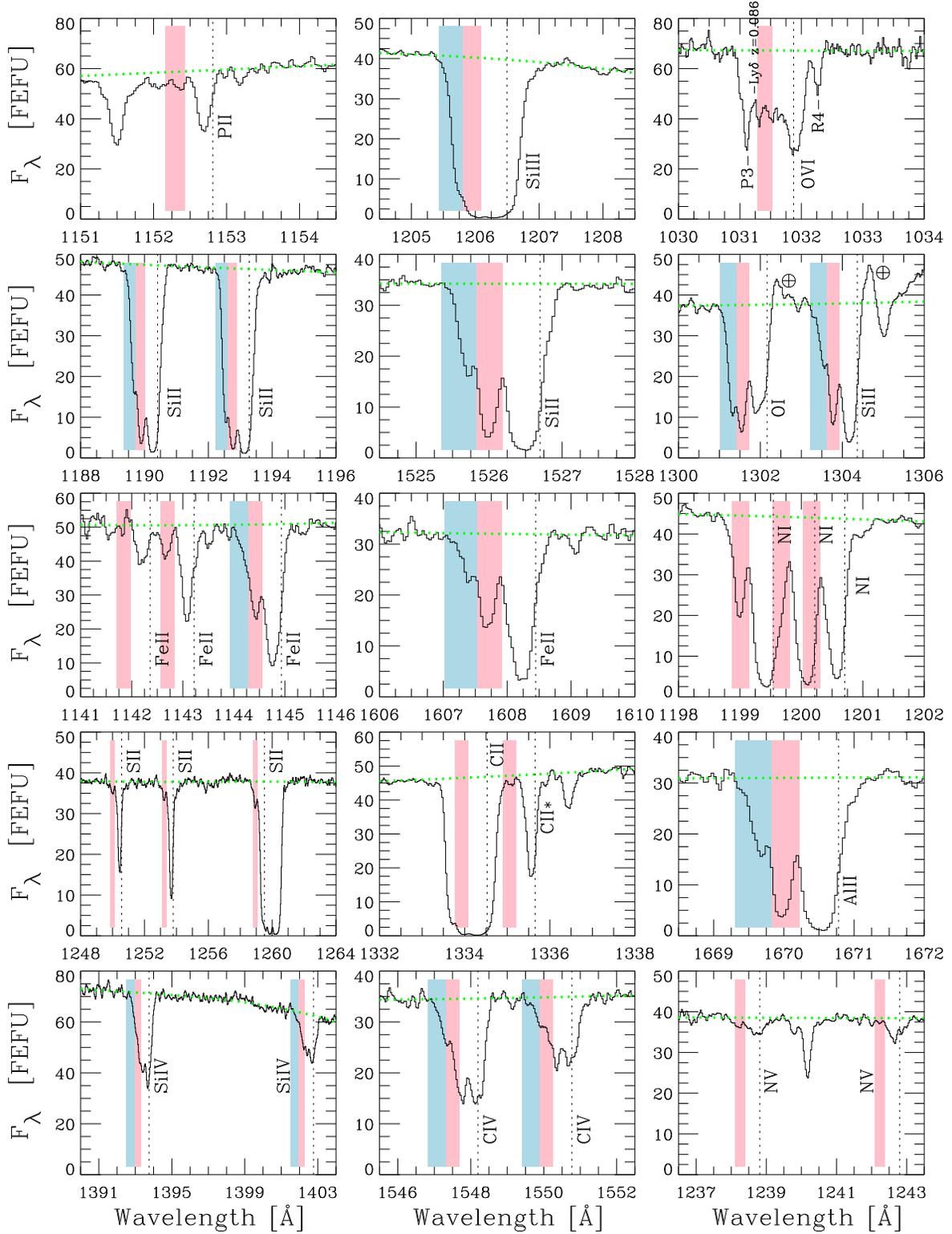}    
   \caption{Same as Fig.\ 4 for Mrk~876 sight line.  Note two HVCs shown by pink and blue wash,
    corresponding to components at $\langle V_{\rm LSR} \rangle = -133$ \kms\ (range $-170$ to
    $-100$ \kms) and $\langle V_{\rm LSR} \rangle = -190$ \kms\ (range $-265$ to $-160$ \kms).
      }
\end{figure} 
  
%%%%%%%%%%%%%%%%%%%%%%%%%%%%%%%%%%%%%%%%%%%%%%%%

%%%%%%%%%%%%%%%%%%%%%%%%%%%%%%%%%%%%%%%%%%%%%%%%%%
%% Figure 7

\begin{figure}
   \epsscale{0.9} 
   \plotone{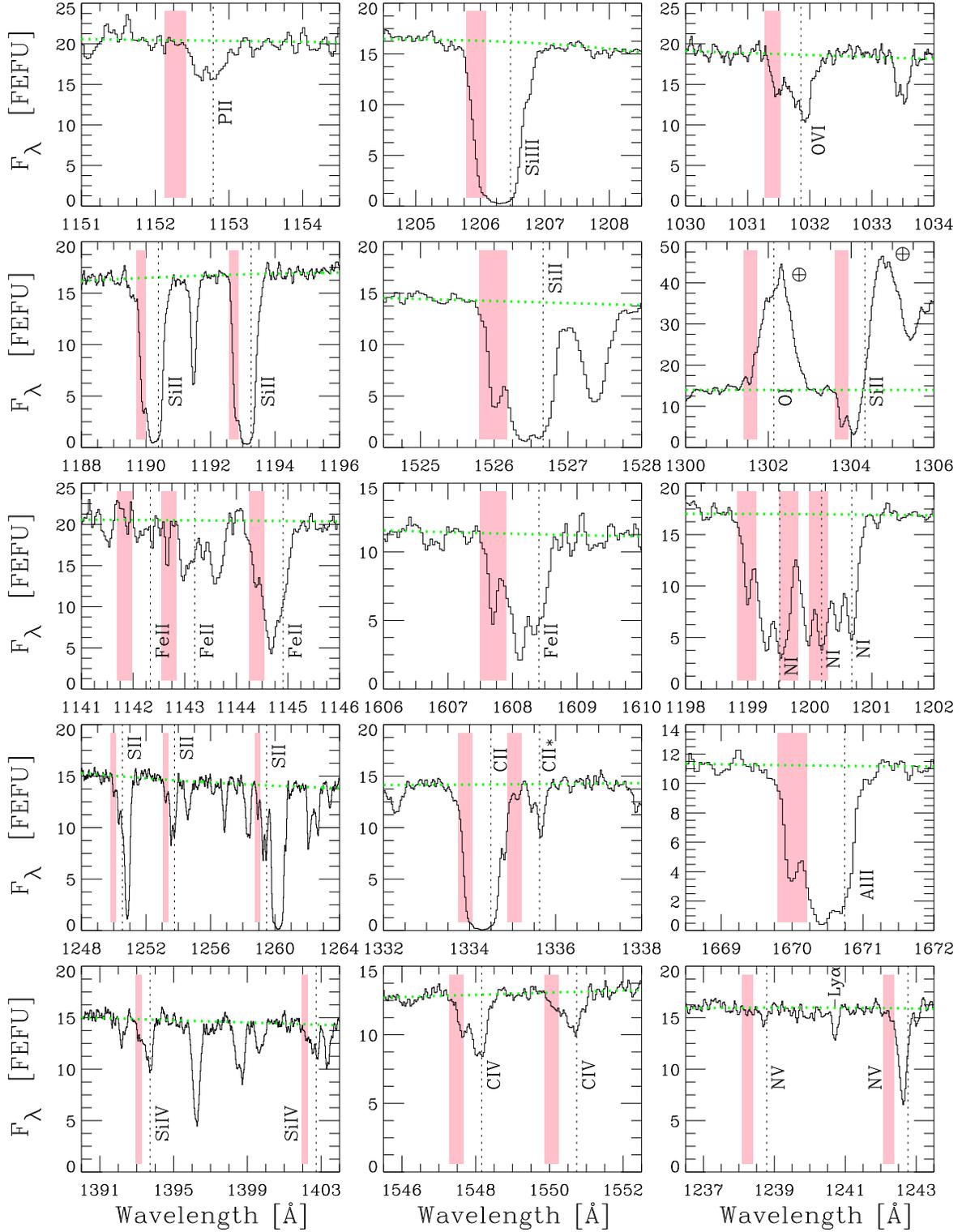}    
   \caption{Same as Fig.\ 4 for PG~1259+593 sight line.   Complex C  (pink wash) is seen at 
      $\langle V_{\rm LSR} \rangle = -128$ \kms\ (range $-170$ to $-95$ \kms). 
      }
\end{figure} 

%%%%%%%%%%%%%%%%%%%%%%%%%%%%%%%%%%%%%%%%%%%%%%%%

%%%%%%%%%%%%%%%%%%%%%%%%%%%%%%%%%%%%%%%%%%%%%%%%%%
%% Figure 8

\begin{figure}
   \epsscale{0.9} 
   \plotone{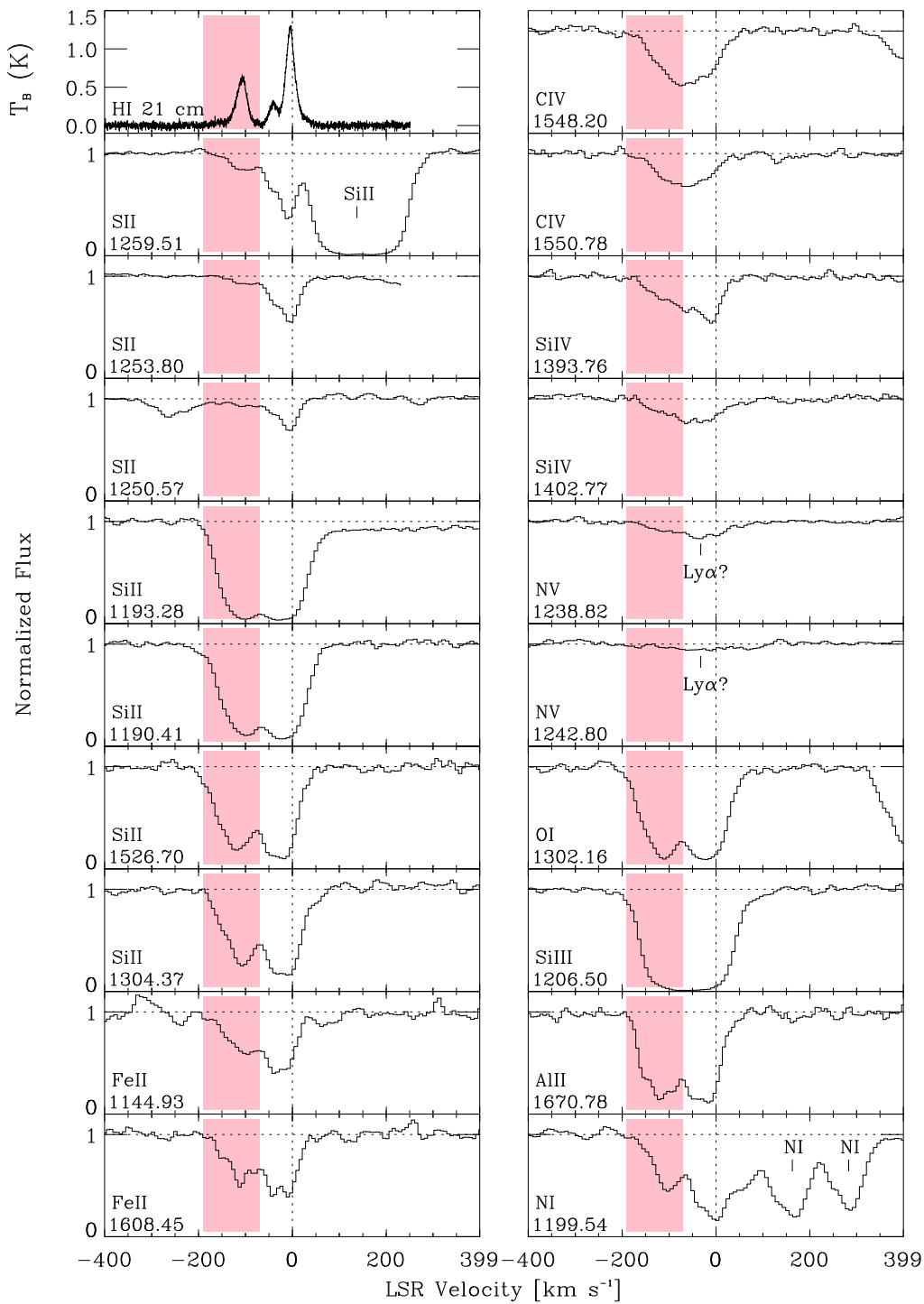}    
   \caption{Plot of  \HI\ emission (top left panel) and UV ion absorption profiles, continuum-normalized, 
       stacked, and aligned in velocity space for the HVC toward Mrk~817.  Complex~C (pink wash) 
       appears between $V_{\rm LSR} = -190$ and $-70$ \kms.  IGM absorbers (\Lya) appear to 
        contaminate both \NV\ lines, which we treat as upper limits. 
      }
\end{figure} 
  
%%%%%%%%%%%%%%%%%%%%%%%%%%%%%%%%%%%%%%%%%%%%%%%%

%%%%%%%%%%%%%%%%%%%%%%%%%%%%%%%%%%%%%%%%%%%%%%%%%%
%% Figure 9

\begin{figure}
   \epsscale{0.9} 
   \plotone{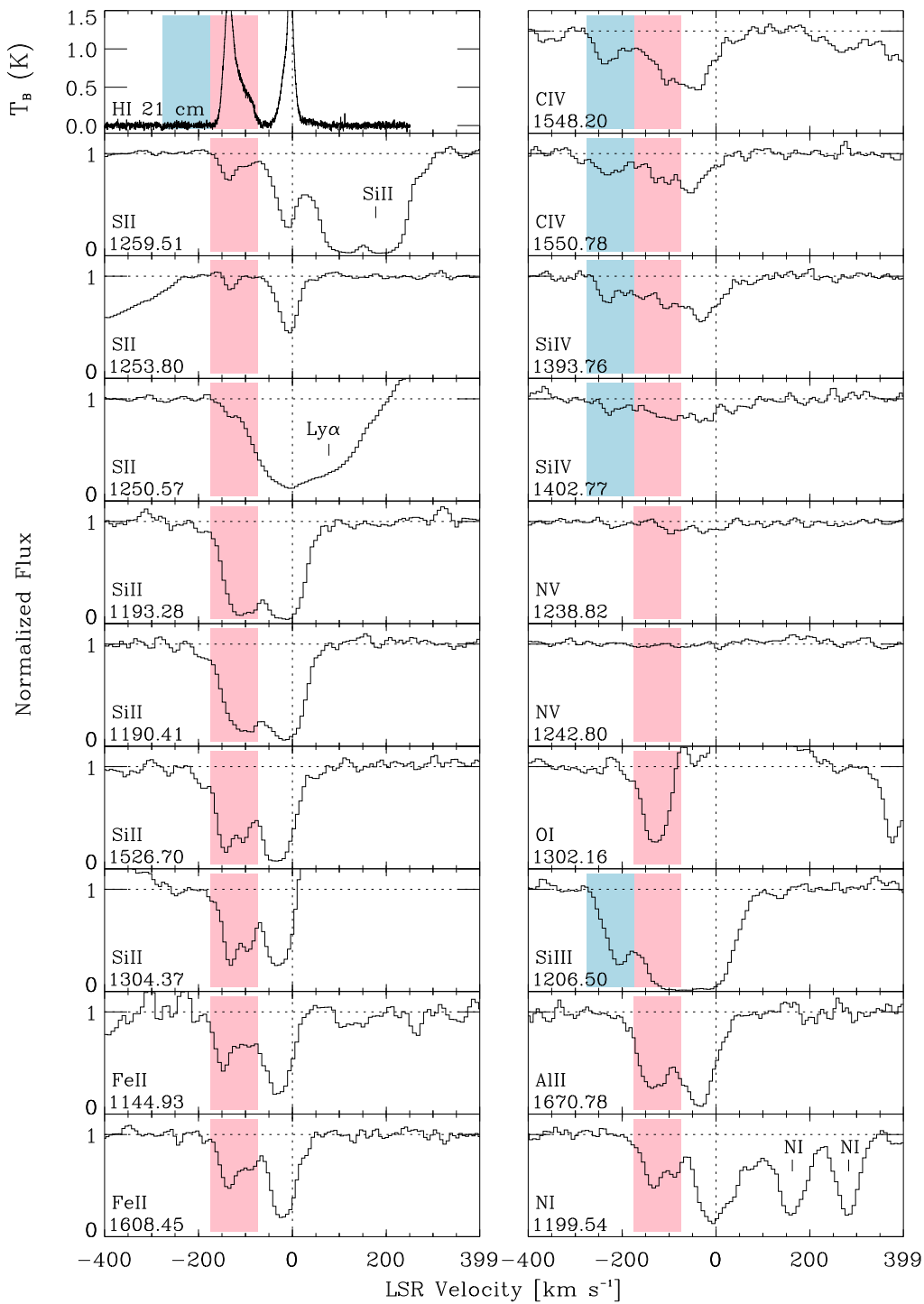}    
   \caption{Same as Fig.\ 8 for two HVCs toward Mrk~290, at  $\langle V_{\rm LSR} \rangle = -120$ \kms\ 
      (pink wash) and $-220$ \kms\ (blue wash). Complex~C lies at $V_{\rm LSR} = -160$ to $-75$ \kms.  
      An IGM absorber (\Lya) is noted in the red wing of \SII\ $\lambda1250$. 
      }
\end{figure} 
  
%%%%%%%%%%%%%%%%%%%%%%%%%%%%%%%%%%%%%%%%%%%%%%%%

%%%%%%%%%%%%%%%%%%%%%%%%%%%%%%%%%%%%%%%%%%%%%%%%%%
%% Figure 10

\begin{figure}
   \epsscale{0.9} 
   \plotone{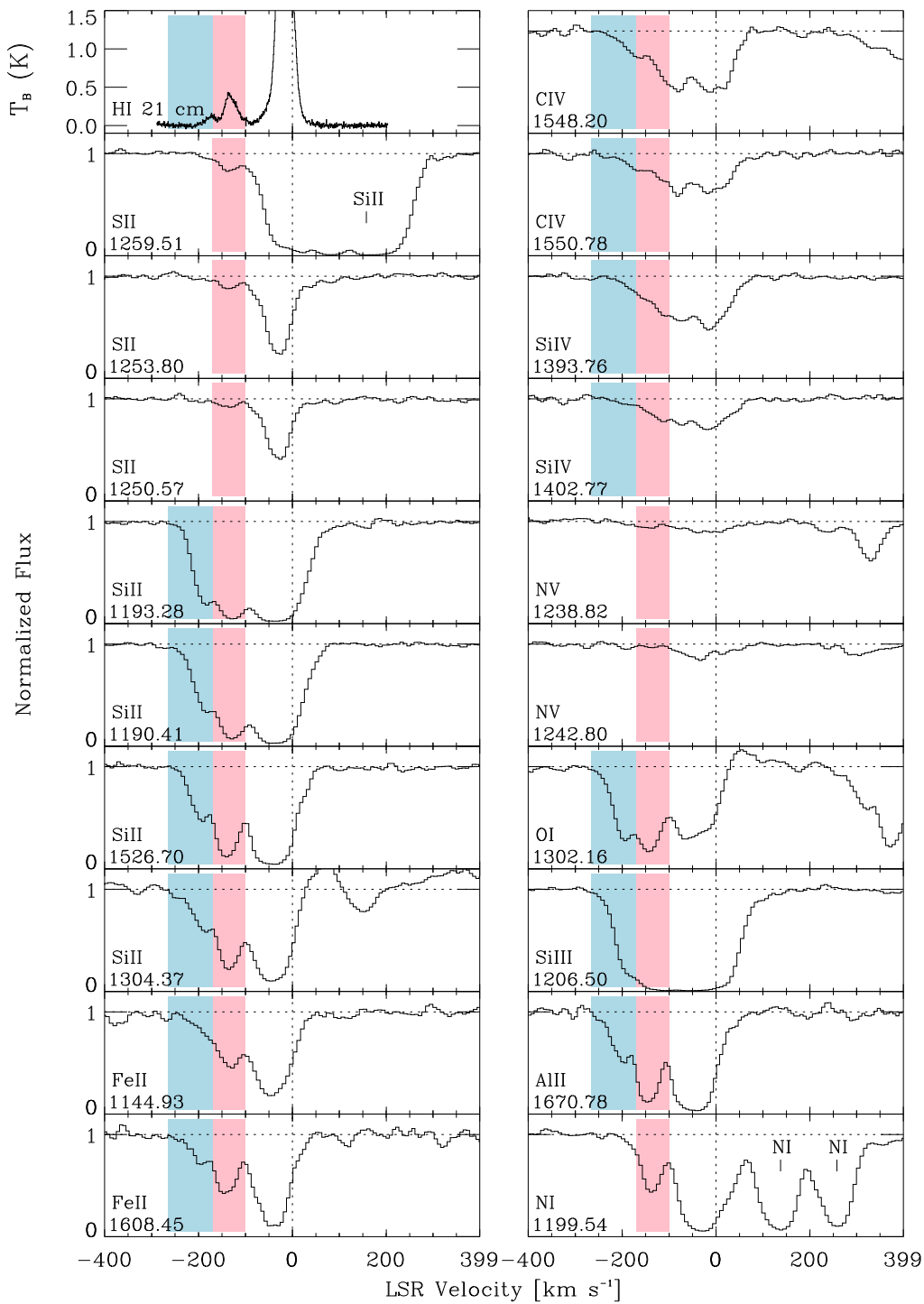}    
   \caption{Same as Fig.\ 8 for HVC toward Mrk~876.  We find two HVCs 
       at $\langle V_{\rm LSR} \rangle = -133$ \kms\ (pink wash) and  
       $\langle V_{\rm LSR} \rangle = -190$ \kms\ (blue wash).
   }
\end{figure} 
  
%%%%%%%%%%%%%%%%%%%%%%%%%%%%%%%%%%%%%%%%%%%%%%%%

%%%%%%%%%%%%%%%%%%%%%%%%%%%%%%%%%%%%%%%%%%%%%%%%%%
%% Figure 11

\begin{figure}
   \epsscale{0.9} 
   \plotone{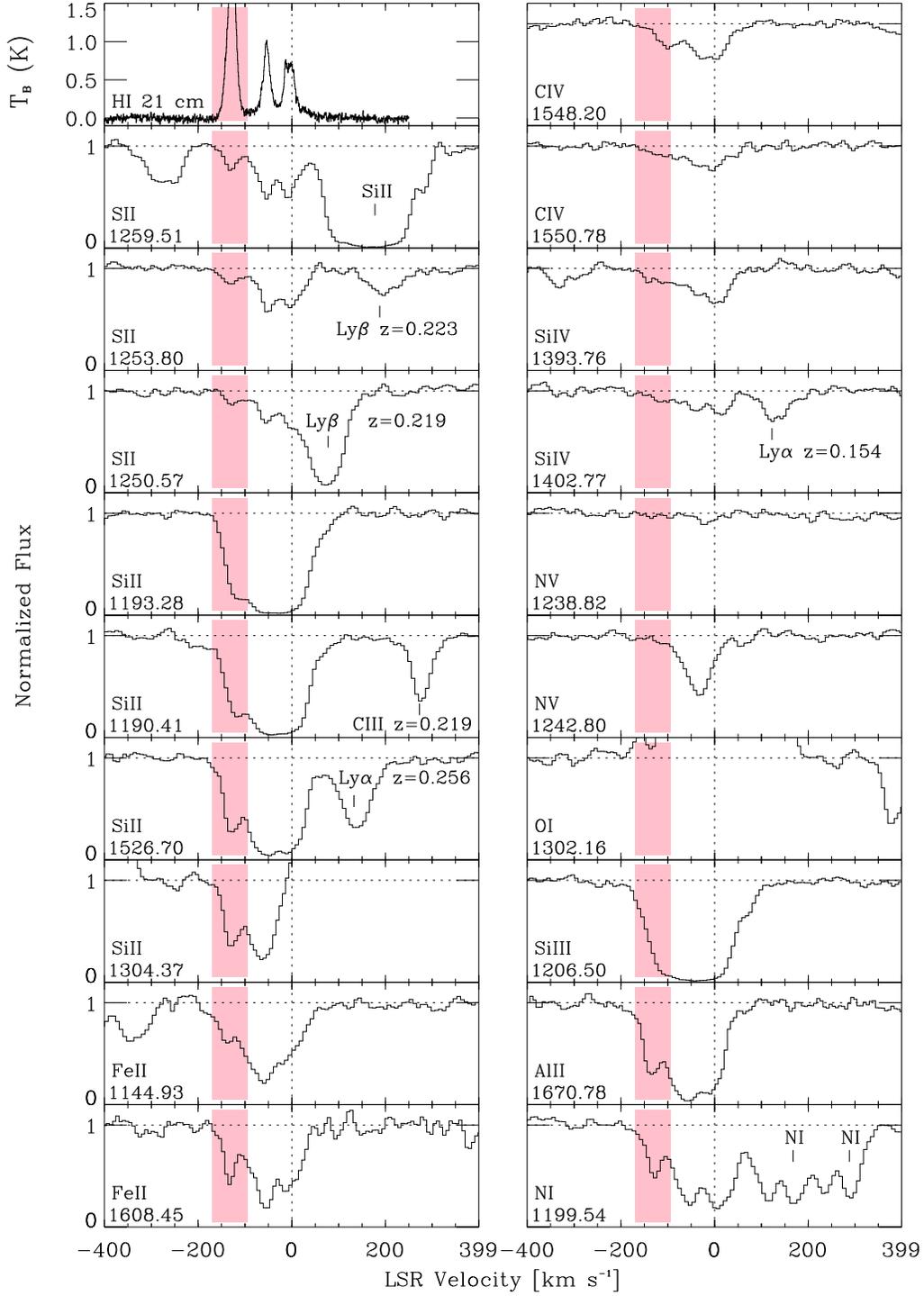}    
   \caption{Same as Fig.\ 8 for HVC at $\langle V_{\rm LSR} \rangle = -128$ \kms\ (pink wash) 
    toward PG~1259+593.  IGM absorbers (\Lya, \Lyb, \CIII) are noted.  
   }
\end{figure} 
  
%%%%%%%%%%%%%%%%%%%%%%%%%%%%%%%%%%%%%%%%%%%%%%%%

\end{document}